\documentclass[twocolumn,showpacs,showkeys,superscriptaddress]{revtex4}
\usepackage{amsmath}
\usepackage{amssymb}
\usepackage{graphicx}

\newcommand{\bastar}{\begin{eqnarray*}}
\newcommand{\eastar}{\end{eqnarray*}}
\newskip\humongous \humongous=0pt plus 1000pt minus 1000pt

\relax
\newcommand{\bea}{\begin{eqnarray}}
\newcommand{\eea}{\end{eqnarray}}
\newcommand{\X}{{\vec X}}
\newcommand{\pro}{\partial}
\newcommand{\hn}{\hat n}
\newcommand{\hr}{\hat r}
\newcommand{\pd}{\partial}
\newcommand{\mn}{{\mu \nu}}
\newcommand{\ab}{{\alpha \beta}}
\newcommand{\hD}{{\hat D}}

\newcommand{\n}{\vec n}

\newcommand{\A}{{\vec A}}

\newcommand{\Int}{{\displaystyle \int}}

\newcommand{\nn}{\nonumber}
\newcommand{\om}{\omega}
\begin{document}

\title{Knots in Physics}
\author{Y. M. Cho}
\email{ymcho0416@gmail.com}
\affiliation{Institute of Modern Physics,
Chinese Academy of Science, Lanzhou 730000, China}  
\affiliation{Center for Quantum Spacetime, Sogang University,
Seoul 04107, Korea}  
\affiliation{School of Physics and Astronomy,
Seoul National University, Seoul 08826, Korea}
\author{Seung Hun Oh}
\affiliation{Center for Quantum Spacetime, Sogang University,
Seoul 04107, Korea}
\affiliation{Department of Physics, Konkuk University, Seoul 05029, Korea}
\author{Pengming Zhang}
\affiliation{Institute of Modern Physics, Chinese Academy of
Science, Lanzhou 730000, China}

\begin{abstract}
~~~~~After Dirac introduced the monopole, topological
objects have played increasingly important roles in physics. 
In this review we discuss the role of the knot, the most 
sophisticated topological object in physics, and related 
topological objects in various areas in physics. In particular, 
we discuss how the knots appear in Maxwell's theory, 
Skyrme theory, and multi-component condensed matter 
physics. 
\end{abstract}
\maketitle

\section{Introduction}

The topological objects have been assuming increasingly important 
roles in physics. The idea of topologically stable matter has first 
been proposed by Lord Kelvin more than a century ago, who suggested 
in 1868 that the atoms could be knots or links of vorticity lines 
of aether \cite{kelvin}. When the topological stability of the knot 
was not well known in physics he understood that such knots and 
links would be extremely stable, just as matter is. This was a remarkable combination of geometric insight and physical intuition. The Kelvin's 
idea was well received at that time, and praised (among others) by 
Maxwell.

Later in 1931, Dirac introduced one of the most important topological 
objects in physics, the Dirac monopole, which has the $\pi_1(S^1)$ 
topology, the first homotopy of $S^1$ \cite{dirac}. Based on this 
topology he proposed that the electric and magnetic charges $e$ 
and $g$ must satisfy the famous Dirac quantization rule $eg=2\pi n$, 
which explaines why the electric charge in nature has discrete values. 
Doing this he demonstrated, for the first time in human history, 
that topology can actually play a fundamental role in nature. 

It was a remarkable coincidence that in the same year Hopf 
has shown that $S^3$ has the so-called Hopf fiberation 
$S^2\times S^1$, and that topologically $\pi_3(S^2)$, 
the third homoptpy of $S^2$, can be classified by the Hopf 
index \cite{hopf,bott}. Mathematically this defines the knot 
topology and the knot number, which will play a crucial role 
in our discussion. 

The Dirac monopole has changed the physics completely, and 
put the topology firmly in physics forever. Since then the monopole 
has become an obsession, theoretically and experimentally. 
Theoretically the non-Abelian monopoles based on $\pi_2(S^2)$ 
topology, the singular Wu-Yang monopole in QCD \cite{wu,prl80} 
and the regularized 'tHooft-Polyakov monopole in spontaneously 
broken gauge theory \cite{thooft,dokos} have been discovered. 
Experimentally huge efforts to discover the monopole have 
been made \cite{cab}. But so far the search for the monopole 
has been unsuccessful, because the monopole which could 
exist in nature is not one of these monopoles but the electroweak 
monopole which is an hybrid between Dirac monopole and 
'tHooft-Polyakov monopole \cite{plb97,yang}. 

In the mean time other interesting topological objects and 
their impacts in physics have become known. The Aharonov-Bohm 
effect and Berry phase are the well known examples of the impact 
of topology in physics \cite{bohm,berry}. But in 1961 Skyrme 
has pushed the topology to a new stage in physics \cite{sky}. 
In his Skyrme theory he tried to realize the Kelvin's dream 
proposing the proton (in general nuclei) to be the skyrmion, 
a topological soliton which has the $\pi_3(S^3)$ topology. 
What is really remarkable is that he was proposing that we 
could construct fermions from bosons with the spin-isospin 
transmutation.

This view has become very successful from the theoretical point
of view \cite{witt,prep}. A systematic method to construct 
multi-skyrmions with large baryon number based on the rational 
map was developed, which has allowed people to construct skyrmions 
with the baryon number up to 108 \cite{man,bat,sut,sky108}.  
With this these multi-skyrmions have been identified with real 
nuclei.  

In particular, the Skyrme theory could explain the spectrum 
of rotational excitations of carbon-12 and the Hoyle state 
successfully, which we need to generate heavy nuclear 
elements in early universe \cite{hoyle,epel}. In spite of 
these sucesses, the Skyrme's proposal was too ideal to 
describe the real world. 

Although the Skyrme theory was originally proposed as 
an effective theory for strong interaction, it has multiple 
faces which allow us to have totally different interpretations. 
First, it has many topologically interesting features and 
has all topological objects known in physics. In addition 
to the well known skyrmion, it has the (helical) baby 
skyrmion and the prototype knot known as the Faddeev-Niemi 
knot \cite{fadd,prl01}. Most importantly, it has the monopole 
which plays the fundamental role. In fact the theory can 
be viewed as a theory of monopole which has a built-in 
Meissner effect, so that all finite energy topological objects 
in the theory could be viewed either as dressed monopoles 
or confined magnetic flux of the monopole-antimonopole 
pair \cite{prl01,plb04,ijmpa08}. 

Of course, the fact that the skyrmion is closely related to 
the monopole has been noticed by many people. In spite of 
the fact that the skyrmions has been classified by the baryon 
number, people have used the rational map $\pi_2(S^2)$ which 
describes the monopole number, not $\pi_3(S^3)$ which 
describes the baryon number, to construct the multi-skyrmion 
solutions \cite{man}. This strongly implies that the skyrmions 
can actually be viewed as the monopoles \cite{prl01,plb04,ijmpa08}. 
For this reason it has been proposed that the skyrmions carry 
two topological numbers, the baryon number and the monopole 
number \cite{epjc17}. 

Furthermore, we can show that the Skyrme theory in fact has 
the multiple vacua which has the $\pi_1(S^1)$ topology 
of the vacuum of Sine-Gordon theory and the $\pi_3(S^2)$ 
topology of QCD vacuum combined together \cite{epjc17}. 
This means that the vacuum of the Skyrme theory can also 
be classified by two integers $(p,q)$, $p$ which represents 
the $\pi_1(S^1)$ topology of the Sine-Gordon vacuum and 
$q$ which represents the $\pi_3(S^2)$ topology of QCD. 
These new features demonstrates that the Skyrme theory has 
more topological features than we thought to have.

But what makes the Skyrme theory very important for the following 
discussion is that it can be reduced to the Skyrme-Faddeev 
theory which describes the prototype knot \cite{prl01,ijmpa08}.
Moreover, we can derive the Skyrme-Faddeev theory which 
describes the core dynamics of the Skyrme theory directly from 
QCD. 

Furthermore, we can argue that, with a simple modification 
the Skyrme theory could describe multi-gap superconductor or
multi-component Bose-Einstein condensate which could contain 
the knot. This strongly indicates that the Skyrme theory 
could also be interpreted to describe a very interesting 
low energy condensed matter physics in a completely different 
environment. This puts the Skyrme theory in a totally new 
perspective.

It has long been believed that we need a complicated theory,
in particular the non-Abelian structure or non-linear structure,
to obtain the knot. This popular wisdom turned out to be totally 
wrong. In physics the knot was first discovered in Maxwell's 
theory, a most unlikely place. In a series of paper in 1989 
Ranada has shown that pure Maxwell's theory without any 
electromagnetic source admits a wide class of propagating 
solitonic knot solutions he called ``pseudo-photons," generalizing 
Trautman's work \cite{ran,nat,knotprep,traut}. This was surprizing. 
 
Introducing two (electric and magnetic) complex scalars which 
is well defined at infinity, he constructed a set of propagating  
electromanetic knot solitons charactrized by two Hopf indices, 
the electric and magnetic helicity, which describe the Hopf map 
from the compactified space $S^3$ to the target space $S^2$. 
Moreover, he showed that this set of knot solutions is dense 
in the sense that a superposition of these topological solitions 
can describe, at least locally, any electromagnetic wave.  

As importantly, in these set he showed that there is a natural 
mechanism for the quantization of electric charge by which 
the electric flux through any closed surface surrounding 
a singularity representing the electric charge becomes 
an integer. If this is true, the electromagnetic knots provides 
a new explanation of the topological charge quantization
which is seemingly independent of the Dirac quantization
rule. It would be very interesting to show if there is any 
relation between the two topological explanations of 
the charge quantization.

In his ``daring conservatism", Wheeler has conjectured 
the existence of ``geons", in particular the electromagnetic 
and gravitational geons, to propose that the elementary 
particles actually could be constructed from electromagnetic 
and/or gravitational fields and thus may not be so 
``elementary" \cite{wheel}. But he could not show 
the existence of such geons. Ranada's electromagnetic knots 
are the perfect examples of such geons. In particular, 
these knots acquire descrete masses in the absence of any 
mass scale in Maxwell's theory, and thus demonstrate that 
the mass can be created from nothing. 

This is really remarkable and astonishing. In QCD the mass 
can be created by the monopole condensation which generates 
the confinement \cite{prd13,prd15}. This is the dimensional 
transmutation in QCD. But here we have another example 
of dimensional transmutation, the mass from the energy, 
without any condensation or any new mechanism. This was 
what Wheeler visioned, and Ranada showed that this is 
possible. And at the center of this is the knot topology.       
It is the topology which makes it possible.

Mathematically the knot is defined as the mapping of a closed 
curve $S^1$ to a three-dimensinal space. The importance 
of this mapping is that this embedding can have non-trivial 
linking. Clearly this linking number does not change in any 
smooth deformation of the curve(s) and thus becomes 
a topological invariant. And in physics we often deal with 
curves; velocity curve, electric current, and magnetic flux. 
When they form closed curves, they may have non-vanishing 
linking number which does not change (i.e., is conserved) 
in dynamical evolotion, because it is a topological invariant. 
This is how the knot becomes important in physics.

Obviously this topological conservation applies to other 
topological objects as well, the magnetic monopoles and 
the skyrmions. And these days we can hardly talk about physics 
without mentioning these topological objects. Many of these 
topological objects are intimately related, but among them 
a most interesting object, mathematically as well as physically, 
is the knot \cite{hopf,atiyah}. And the knot appears everywhere 
in physics; in Maxwell's electrodynamics \cite{ran,berry1}, 
Skyrme theory \cite{prl01,plb04,ijmpa08}, QCD \cite{plb05},   
fluid dynamics \cite{moff}, atomic physics \cite{berry2,pra05}, 
plasma physics \cite{berg}, polymer physics \cite{kami}, 
condensed matter physics \cite{baba,prb06,epjb08}, even in 
Einstein's theory \cite{cqg13}. 

The purpose of this paper is to review the role of the knots 
and related topological objects, in particular the monopoles 
and skyrmions, in physics and to discuss their importance.
Although the knot appears in all area of physics, we will 
discuss the knot in Skyrme theory in detail because the Skyrme 
theory is a best place to discuss the topological objects in physics.    

The paper is organized as follows. In Section II we discuss 
the Ranada's construction of the electromagnetic knot and 
its physical implication. In Section III we review the Skyrme 
theory, and show that the core dynamics of the theory 
is described by the Skyrme-Faddeev theory which has 
the non-Abelian monopole and the prototype knot. In 
Section IV we show that the Skyrme-Faddeev theory can 
actually be derived from the SU(2) QCD, and discuss a deep 
connection between the two theories. In Section V we discuss 
two types of topology in skyrmions, and show that skyrmions 
actually carry two different topology, the $\pi_2(S^2)$ 
topology of the monopole as well as the well known 
$\pi_3(S^3)$ topology of skyrmion. In Section VI we discuss 
the vacuum structure of Skyrme theory, and show that it 
has the multiple vacua made of the vacuum topology of 
Sine-Gordon theory and QCD combined together. In Section 
VII we review the prototype knot in Skyrme theory which 
can be viewed as the twisted magnetic vortex ring made of 
the monopole-antimonopole pair. In Section VIII we show 
how similar knots could appear in condensed matter physics, 
in particular in multi-gap superconductors and multi-component 
Bose-Einstein condensate. Finally in the last section we discuss 
the physical implications of the knots. 

\section{Knots in Maxwell's Theory}

Maxwell's theory has played fundamental role in physics. 
In spite of the mathematical beauty and the huge practical 
applications, however, the theory has been thought to be 
too simple to admit topological objects. Of course, Dirac 
showed us that it can be generalized to allow the U(1) 
monopole structure, but as far as topology is concerned 
this was all. Fortunately this general wisdom turned out 
to be completely wrong.

In fact the Maxwell's theory, without Dirac's generalization, 
has very interesting knots \cite{ran}. But this important 
topological feature of Maxwell's theory so far has not 
been well known, because people believed that the theory, 
being linear, has no structure for topological object. So 
we discuss the knots in Maxwell's theory first.

To understand how this could be possible, notice that 
the electric and magnetic field lines could form closed 
loops and thus be linked, and this linking could provide 
the topological structure. Let two closed field lines be 
$\vec c_1(s)$ and $\vec c_2(s)$. They are linked if they 
have a non-vanishing Gauss integral,
\bea
L(\vec c_1,\vec c_2)=\dfrac1{4\pi} \Int  
\Big(\dfrac{\vec c_1-\vec c_2}{|\vec c_1-\vec c_2|^3}
\times \dfrac{\vec c_1}{ds_1} \Big)
\cdot \dfrac{\vec c_2}{ds_2} ds_1 ds_2,
\eea
where the self-linking number $L(\vec c,\vec c)$ describes 
the knottedness. In the case of the electric and magnetic 
fields the average of the the linking integral over all field 
line pairs together with the self-linking number over all 
field lines give rise to the electric and magnetic helicities 
given by the Chern-Simon integrals, 
\bea
&h_e=\Int \epsilon_{ijk} C_i (\pd_j C_j-\pd_k C_j)~d^3x,  \nn\\
&h_m=\Int \epsilon_{ijk} A_i (\pd_j A_j-\pd_k A_j)~d^3x.
\label{hel}
\eea
Here $C_i$ and $A_i$ are the vector potentials of the electric
and magnetic field, which exist in free Maxwell's theory. 
And in Maxwell's theory these integer helicities determine 
the $\pi_3(S^2)$ topology of the electromagnetic knots. 

To explain this in more detail we first construct the electromagnetic 
knot. Let $\zeta$ and $\eta$ be two complex scalar fields 
representing the map from the space-time $R^3\times T$ 
to the complex plane $R^2$. Compactifying $R^3$ and $R^2$ 
to $S^3$ and $S^2$ with the streographic projection, we 
can view that the two complex fields define the map from 
$S^3$ to $S^2$ which can be classified by the knot topology 
$\pi_3(S^2)$.   

Next, we define two anti-symmetric electromagnetic tensor 
fields $F_\mn$ and $H_\mn$ by
\bea
&F_\mn= \dfrac{\sqrt \alpha}{2\pi i} 
\times \dfrac{\pd_\mu \zeta^* \pd_\nu \zeta
-\pd_\nu \zeta^* \pd_\mu \zeta}{(1+\zeta^* \zeta)^2},   \nn\\
&H_\mn=\dfrac{\sqrt \alpha}{2\pi i} \times 
\dfrac{\pd_\mu \eta^* \pd_\nu \eta
-\pd_\nu \eta^* \pd_\mu \eta}{(1+\eta^* \eta)^2},
\label{def}
\eea
where we have put an action constant $\alpha$ to make
the tensor fields to have the dimension of the electromagnetic 
fields. We further require $F_\mn$ and $H_\mn$ to be dual 
to each other,
\bea
&F_\mn=\dfrac12 \epsilon_{\mn \ab} H^\ab,
~~~H_\mn=-\dfrac12 \epsilon_{\mn \ab} F^\ab.
\label{dual}
\eea
Now we can prove that, if $\zeta$ and $\eta$ satisfy 
the equation
\bea
&\pd_\mu F^{\mn} \pd_\nu \zeta=0,
~~~\pd_\mu F^{\mn} \pd_\nu \zeta^*=0,  \nn\\
&\pd_\mu H^{\mn} \pd_\nu \eta=0,
~~~\pd_\mu H^{\mn} \pd_\nu \eta^*=0,
\label{eom}
\eea 
$F_\mn$ and $H_\mn$ become the solutions of Maxwell's 
equation, and thus describe the electromagnetic 
fields \cite{ran}.

Notice that (\ref{eom}) tells that, if the Cauchy data 
$(\zeta,\dot \zeta,\eta,\dot \eta)$ at $t=0$ satisfies 
the duality condition (\ref{dual}), it does so for all $t$.
Moreover, (\ref{def}) assures that $F_\mn$ and $H_\mn$
satisfy the first half of Maxwell's equation
\bea
\epsilon_{\mn\ab} \pd^\nu F^\ab=0,
~~~\epsilon_{\mn\ab} \pd^\nu H^\ab=0,
\eea
independent of $\zeta$ and $\eta$. Besides, this together
with the duality condition (\ref{dual}) assures that they 
satisfy the second half of Maxwell's equation
\bea
\pd_\mu F^\mn=0,~~~\pd_\mu H^\mn=0.
\eea
So $F_\mn$ and $H_\mn$ become solutions of Maxwell's 
theory. The important point here is that these solutions 
are encoded by the electric and magnetic helicities of 
the $\pi_3(S^2)$ topology given by $\zeta$ and $\eta$. 

A few comments are in order. First, (\ref{def}) tells that
$\vec E(\zeta)$ and $\vec B(\zeta)$ are mutually orthogonal.
Moreover, $\vec B(\zeta)$ and $\vec B(\eta)$ are tangent 
to the curve $\zeta=const$ and $\eta=const$. So the two 
fields $\vec E$ and $\vec B$ are orthogonal to each other
($\vec E \cdot \vec B=0$), and the solutions represent
the radiational wave. Second, although we have introduced 
two complex scalar fields $\zeta$ and $\eta$, it must be 
clear that only one is indepent. This is because $F_\mn$ 
and $H_\mn$ are dual to each other. 

To find the solutions, we have to solve (\ref{eom}). This is 
a non-trivial task. So, in stead of solving the equations, 
Ranada constructed what he called an admissible set of 
solutions classified by two Hoph numbers, the electric 
and magnetic helicities, by judiciously choosing $\zeta$ 
which describes the Hopf map. 

Consider the map proposed by Hopf \cite{hopf}
\begin{gather}
\phi_H=\frac{2(x+iy)}{2z+i(r^2-1)},
~~~(r^2=x^2+y^2+z^2).
\end{gather}
Notice that any two level curves of this mapping, for example 
the curve $\phi_H=0$ and the curve $\phi_H=\infty$, are linked 
once. So the mapping defines the $\pi_3(S^2)$ map with
Hopf index one. Now, let us define $\zeta$ and $\eta$ by
\begin{gather}
\zeta=\phi_H(\kappa y,\kappa z, \kappa x),
~~~\eta=\phi_H^*(\kappa z,\kappa x, \kappa y),
\end{gather}
where $\kappa$ is a dimensional parameter with dimension 
of mass (inverse length). Then we can easily check 
the electromagnetic field given by (\ref{def}) satisfies 
the Maxwell's equation which has $h_e=h_m=1$. This solution 
is called the electromagnetic knot \cite{ran}.

The energy, momentum, and angular momentum of this solution
can be computed from the energy density, momentum density, 
and angular momentum density, $(\vec E^2+ \vec B^2)/2$, 
$\vec E\times \vec B$, and $\vec x \times (\vec E\times \vec B)$.
They are given by
\begin{gather}
E=2\kappa,~~~\vec p=(0,0,\kappa),~~~\vec J=(0,0,1).
\end{gather}
The solution describes a wavepacket given by the potential 
(with $A_0=0$)
\begin{gather}
\vec A= \frac{\kappa^2}{(2\pi)^{3/2}} 
\Int \frac1\omega \Big[\vec R_2(\vec k) 
\cos(\vec k\cdot \vec r-\omega t) \nn\\
+\vec R_1(\vec k)\sin(\vec k\cdot \vec r-\omega t) \Big] d^3 k,  \nn\\
\vec R_1=\frac{\exp(-\omega/\kappa)}{(2\pi)^{1/2} \kappa}
\Big(-\frac{k_1 k_3}{\omega},\frac{k_2^2+k_3^2}{\omega}+k_2,    \nn\\
-\frac{k_1 k_2}{\omega}-k_1 \Big),  \nn\\
\vec R_2=\frac{\exp(-\omega/\kappa)}{(2\pi)^{1/2} \kappa} 
\Big(-\frac{k_1^2+k_2^2}{\omega}-k_2,   \nn\\
\frac{k_1 k_3}{\omega}, \frac{k_2 k_3}{\omega}+k_3 \Big), 
\end{gather}
traveling along the $z$-axis with velocity $p/E=1/2$
which has the electromagnetic field given by
\begin{gather}
\vec E=\frac{\kappa^2}{(2\pi)^{3/2}} 
\Int \Big[\vec R_1(\vec k)\cos(\vec k\cdot \vec r-\omega t) \nn\\
-\vec R_2(\vec k)\sin(\vec k\cdot \vec r-\omega t) \Big] d^3 k,  \nn\\
\vec B=\frac{\kappa^2}{(2\pi)^{3/2}} 
\Int \Big[\vec R_2(\vec k)\cos(\vec k\cdot \vec r-\omega t)  \nn\\
+\vec R_1(\vec k)\sin(\vec k\cdot \vec r-\omega t) \Big] d^3 k.
\end{gather}  
Notice that this wavepacket has the mass 
$m^2= E^2-{\vec p}^2=3\kappa^2$. Clearly this is 
the electromagnetic geon visioned by Wheeler. The fact
that we have such topologically stable finite energy 
solitonic solutions in Maxwell's theory is really remarkable. 

Ranada showed that one can construct a whole set of knot 
solutions with arbitrary integer helicities \cite{ran}. 
Moreover, he showed that this set is dense, in the sense 
that we can construct any radiational wave superposing 
them, at least locally. 

What is more remarkable is that in this formalism the electric 
charge has to be quantized. Introducing a point singularity 
which represents the electric charge in these solutions he 
showed that the total charge, i.e., the integral of the outgoing 
electric flux over the closed surface surrounding the singularity, 
must be quantized and have discrete values. This is basically 
because these point singularities aquires the $\pi_2(S^2)$ 
topology which necessitates the electric charge quantization \cite{ran}. 
Notice that here the $\pi_2(S^2)$ topology describes the electric,
not magnetic, charge. Doing so, he has found another reason 
why the electric charge in nature is quantized. This is really 
remarkable because here we do not have to assume the existence 
of the monopole to have the electric charge quantization.   

Certainly the electromagnetic knots are very interesting from 
the theoretical point of view. As importantly, they have huge 
potential applications in many area, from colloidal and atomic 
particle trapping to the plasma confinement \cite{nat}. But we 
have to keep in mind that, unlike other known topological 
objects in physics, the electromagnetic knot has a non-trivial 
time dependence. 
 
\section{Skyrme Theory: A Review}

The Skyrme theory has played an important role in physics.
Originally it was proposed as a theory of pion physics in strong 
interaction where the skyrmion, a topological soliton made of 
pions, appears as the baryon \cite{sky}. Soon after, the theory 
has been interpreted as a low energy effective theory of QCD 
in which the massless pion fields emerge as the Nambu-Goldstone 
boson of the spontaneous chiral symmetry breaking \cite{witt,prep}. 
This view has become popular and very successful, as we have 
already mentioned \cite{man,bat,sut,sky108,hoyle,epel}. 

But the Skyrme theory has multiple faces and has other 
topologically interesting objects. In fact we can say that 
the theory has all topological objects known in physics.
In addition to the well known skyrmion which has 
the $\pi_3(S^3)$ topology it has the baby skyrmion 
which has the $\pi_1(S^1)$ topology, the monopole 
which has the $\pi_2(S^2)$ topology, and the knot
which has the $\pi_3(S^2)$ topology. 

Among these the monopole plays the fundamental role. 
The central piece of the theory is the singular Wu-Yang 
type monopole which makes up all topological objects 
in the theory. Indeed, the skyrmion can be viewed as 
a finite energy dressed monopole, the baby skyrmion 
as a magnetic vortex connecting the monopole-antimonopole 
pair, and the knot as a twisted magnetic vortex ring made 
of the helical baby skyrmion. So the theory can be viewed 
as a theory of monopole which has a built-in Meissner 
effect \cite{prl01,plb04,ijmpa08}.  

Moreover, the Skyrme theory has the multiple vacua similar
to the Sine-Gordon theory, each of which has the knot 
topology of the QCD vacuum \cite{epjc17}. In other words, 
the vacuum of the Skyrme theory has the topology of 
the Sine-Gordon theory and QCD combined together. 
So, the vacuum of the Skyrme theory can also be classified 
by two integers by $(p,q)$, $p$ which represents
the $\pi_1(S^1)$ topology and $q$ which represents 
the $\pi_3(S^2)$ topology.

It has been appreciated for a long time that the skyrmion 
is described by the monopole topology. But since the monopole 
topology $\pi_2(S^2)$ and the baryon topology $\pi_3(S^3)$ 
are different, the skyrmions should be classified by two 
topological numbers, the baryon number $b$ and the monopole 
number $m$ \cite{epjc17}. According to this view the baryon 
number can also be replaced by the radial (shell) quantum 
number $n$ which describes the $\pi_1(S^1)$ topology of 
the radial extension of the skyrmions. This is based on 
the observation that the $S^3$ space (both the compactified 
3-dimensional real space and the SU(2) target space) has 
the Hopf fibering $S^3\simeq S^2\times S^1$, so that 
the baryon number defined by $\pi_3(S^3)$ can be decomposed 
to two topological numbers $\pi_2(S^2)$ and $\pi_1(S^1)$.

The fact that the skyrmions have two topology can be 
extended to the baby skyrmion. We can construct new 
solutions of baby skyrmion, and show that they have 
two topology, the shell (radial) topology $\pi_1(S^1)$ 
and the monopole topology $\pi_2(S^2)$. This is really 
remarkable, which strongly support that the skyrmion 
does carry two topology. These unique features make 
the Skyrme theory an ideal place for us to discuss 
the topology in physics. 

We first review the well known facts in Skyrme theory. 
Let $\omega$ and $\hn$ (${\hn}^2 = 1$) be the massless 
scalar field (the ``sigma-field") and the normalized 
pion-field in Skyrme theory, and write the Skyrme Lagrangian
as \cite{sky}
\bea
&{\cal L} = \dfrac{\kappa^2}{4} {\rm tr} ~L_\mu^2
+ \dfrac{\alpha}{32}{\rm tr}
\left( \left[ L_\mu, L_\nu \right] \right)^2 \nn\\
&= - \dfrac{\kappa^2}{4} \Big[ \dfrac{1}{2} (\pd_\mu \om)^2
+2 \sin^2 \dfrac{\om}{2} (\pd_\mu \hn)^2 \Big]  \nn\\
&-\dfrac{\alpha}{8} \Big[ \sin^2 \dfrac{\om}{2}
\big((\pd_\mu \om)^2 (\pd_\nu \hn)^2
-(\pd_\mu \om \pd_\nu \om)
(\pd_\mu \hn \cdot \pd_\nu \hn) \big)  \nn\\
&+2 \sin^4 \dfrac{\om}{2}
(\pd_\mu \hn \times \pd_\nu \hn)^2 \Big], \nn\\
&L_\mu = U\pd_\mu U^{\dagger}, \nn\\
&U = \exp (\dfrac{\om}{2i} \vec \sigma \cdot \hn)
= \cos \dfrac{\om}{2} - i (\vec \sigma \cdot \hn)
\sin \dfrac{\om}{2},
\label{slag}
\eea
where $\kappa$ and $\alpha$ are the coupling constants.

Notice that, with
\bea
&\sigma=\cos \dfrac{\omega}{2},
~~~\vec \pi= \hn \sin \dfrac{\omega}{2},
~~~(\sigma^2 + \vec \pi^2 = 1),
\eea
the Lagrangian (\ref{slag}) can also be put into the form
\bea
&{\cal L} = -\dfrac{\kappa^2}{2} \big((\partial_\mu \sigma)^2
+(\partial_\mu \vec \pi)^2 \big) \nn\\
&-\dfrac{\alpha}{4} \big((\partial_\mu \sigma \partial_\nu \vec \pi
- \partial_\nu \sigma \partial_\mu \vec \pi)^2
+ (\partial_\mu \vec \pi \times \partial_\nu \vec \pi)^2 \big) \nn\\
&+\dfrac{\lambda}{4} (\sigma^2 + \vec \pi^2 - 1),
\label{smlag}
\eea
where $\lambda$ is a Lagrange multiplier. In this form $\sigma$
and $\vec \pi$ represent the sigma field and the pion field,
so that the theory describes the non-linear sigma model of
the pion physics.

The Lagrangian has a hidden $U(1)$ gauge symmetry
as well as the global $SU(2)_L \times SU(2)_R$
symmetry \cite{plb04,ijmpa08}. The global
$SU(2)_L \times SU(2)_R$ symmetry is obvious, but
the hidden $U(1)$ gauge symmetry is not. The hidden
$U(1)$ gauge symmetry comes from the $U(1)$ subgroup
which leaves $\hn$ invariant. To see this, we
reparametrize $\hn$ by the $CP^1$ field $\xi$,
\bea
\hn = \xi^\dag \vec \sigma \xi,
~~~~~\xi^\dag \xi=1.
\label{ndef}
\eea
and find that under the $U(1)$ gauge transformation
of $\xi$ to
\bea
\xi \rightarrow \exp (i\theta(x)) \xi,
\label{u1}
\eea
$\hn$ (and $\pro_\mu \hn$) remains invariant.
Now, we introduce the composite gauge potential
$C_\mu$ and the covariant derivative $D_\mu$
which transforms gauge covariantly under (\ref{u1})
by
\begin{gather}
C_\mu =-2i \xi^\dagger \pd_\mu \xi,
~~~D_\mu \xi = (\pd_\mu -\frac{i}{2} C_\mu )\xi.
\label{mpot}
\end{gather}
With this we have the following identities,
\begin{gather}
(\pd_\mu \hn )^2 = 4 |D_\mu \xi |^2,  \nn\\
\pd_\mu \hn \times \pd_\nu \hn 
=-2i \Big[(\pd_\mu \xi^\dagger)(\pd_\nu \xi) 
- (\pd_\mu \xi^\dagger)(\pd_\nu\xi)\Big] \hn \nn\\
= H_\mn \hn,  \nn\\
H_\mn=\pd_\mu C_\nu - \pd_\nu C_\mu.
\end{gather}
Furthermore, with the Fierz' identity
\begin{gather}
\sigma_{ij}^a \sigma_{kl}^a = 2\delta_{il}\delta_{jk}
- \delta_{ij}\delta_{kl},
\end{gather}
we have
\begin{gather}
\pd_\mu \hn \cdot \pd_\mu \hn
=2 \pd_\mu (\xi^\dagger_i \xi_j)
\pd_\nu (\xi^\dagger_j \xi_i)  \nn\\
= 2\Big[ (\pd_\mu\xi^\dagger\xi )(\pd_\nu \xi^\dagger\xi)
+(\pd_\mu \xi^\dagger)( \pd_\nu \xi )  
+(\partial_\nu \xi^\dagger)( \partial_\mu \xi)  \nn\\
+(\xi^\dagger \pd_\mu \xi)(\xi^\dagger \pd_\nu \xi) \Big] 
=2\Big[(\pd_\mu \xi^\dagger + \frac{i}{2} C_\mu \xi^\dagger )
(\pd_\nu \xi -\frac{i}{2} C_\nu \xi)   \nn\\
+(\pd_\nu \xi^\dagger + \frac{i}{2} C_\nu \xi^\dagger )
(\pd_\mu \xi - \frac{i}{2} C_\mu \xi) \Big] \nn\\
= 2\Big[ (D_\mu \xi )^\dagger (D_\nu \xi)
+ (D_\nu \xi )^\dagger (D_\mu \xi)\Big].
\end{gather}
From this we can express (\ref{slag}) by
\begin{gather}
{\cal L} = - \dfrac{\kappa^2}{4}
\Big[ \dfrac{1}{2} (\pd_\mu \om)^2
+8 \sin^2 \dfrac{\om}{2} |D_\mu \xi|^2 \Big]  \nn\\
-\dfrac{\alpha}{2}  \sin^2 \dfrac{\om}{2}
\Big[(\pd_\mu \om)^2 |D_\mu \xi|^2
-(\pd_\mu \om \pd_\nu \om)
(D_\mu \xi )^\dagger (D_\nu \xi) \nn\\
+\frac{1}{2}\sin^2 \dfrac{\om}{2} H_\mn^2 \Big],
\label{skyu1}
\end{gather}
which is explicitly invariant under the $U(1)$ gauge
transformation (\ref{u1}). So replacing $\hn$ by
$\xi$ in the Lagrangian we can make the hidden
$U(1)$ gauge symmetry explicit. In this form
the Skyrme theory becomes a self-interacting
$U(1)$ gauge theory of $CP^1$ field coupled to
a massless scalar field.

The equations of motion of the Lagrangian is given by
\begin{gather}
\pd^2 \om -\sin\om (\pd_\mu \hn)^2
+\dfrac{\alpha}{8 \kappa^2} \sin\om (\pd_\mu \om
\pd_\nu \hn -\pd_\nu \om \pd_\mu \hn)^2  \nn\\
+\dfrac{\alpha}{\kappa^2} \sin^2 \dfrac{\om}{2}
\pd_\mu \big[ (\pd_\mu \om \pd_\nu \hn
-\pd_\nu \om \pd_\mu \hn) \cdot \pd_\nu \hn \big] \nn\\
- \dfrac{\alpha}{\kappa^2} \sin^2 \dfrac{\om}{2}
\sin\om (\pd_\mu \hn \times \pd_\nu \hn)^2 =0, \nn \\
\pd_\mu \Big\{\sin^2 \dfrac{\om}{2}  \hn \times
\pd_\mu \hn + \dfrac{\alpha}{4\kappa^2} \sin^2 \dfrac{\om}{2}
\big[ (\pd_\nu \om)^2 \hn \times \pd_\mu \hn  \nn\\
-(\pd_\mu \om \pd_\nu \om) \hn \times \pd_\nu \hn \big] \nn\\
+\dfrac{\alpha}{\kappa^2} \sin^4 \dfrac{\om}{2} (\hn \cdot
\pd_\mu \hn \times \pd_\nu \hn) \pd_\nu \hn \Big\}=0.
\label{skeq1}
\end{gather}
Clearly the second equation describes the conservation of 
$SU(2)$ current originating from the global $SU(2)$ symmetry 
of the theory.

It has two interesting limits. First, when
\bea
\om= (2p+1) \pi,
\label{fadd}
\eea
(\ref{skeq1}) is reduced to
\begin{gather}
\hn \times \pd^2 \hn-\dfrac{\alpha}{\kappa^2} 
(\pd_\mu H_\mn) \pd_\nu \hn = 0, \nn\\
H_\mn=\hn \cdot (\pd_\mu \hn \times \pd_\nu \hn)
= \pd_\mu C_\nu - \pd_\nu C_\mu,
\label{sfeq}
\end{gather}
where $C_\mu$ is the magnetic potential of $H_\mn$ defined
by (\ref{mpot}). This is the central equation of Skyrme theory
which allows the monopole, the baby skyrmion, and
the Faddeev-Niemi knot \cite{prl01,plb04,ijmpa08}.

Second, in the spherically symmetric limit
\bea
\om = \om (r),~~~~~\hn = \pm \hat r,
\label{skans}
\eea
(\ref{skeq1}) is reduced to
\bea
&\dfrac{d^2 \om}{dr^2} +\dfrac{2}{r} \dfrac{d\om}{dr}
-\dfrac{2\sin\om}{r^2} +\dfrac{2\alpha}{\kappa^2}
\Big[\dfrac{\sin^2 (\om/2)}{r^2} \dfrac{d^2 \om}{dr^2}  \nn\\
&+\dfrac{\sin\om}{4 r^2} (\dfrac{d\om}{dr})^2
-\dfrac{\sin\om \sin^2 (\om /2)}{r^4} \Big] =0.
\label{skeq2}
\eea
This is the equation used by Skyrme to find the original
skyrmion. Imposing the boundary
condition
\bea
\om(0)= 2\pi,~~~~~\om(\infty)= 0,
\label{skbc}
\eea
we have the skyrmion solution which carries the unit
baryon number \cite{sky}
\bea
&B=-\dfrac{1}{24\pi^2} \Int
\epsilon_{ijk} ~{\rm tr} ~(L_i L_j L_k) d^3r \nn\\
&= -\dfrac{1}{8\pi^2} \int \epsilon_{ijk} \pd_i \om
\big[\hat r \cdot (\pd_j \hat r \times \pd_k \hat r) \big]
\sin^2 \dfrac{\om}{2} d^3r \nn\\
&=1,
\label{bn}
\eea
which represents the non-trivial homotopy $\pi_3(S^3)$
defined by $U$.

The two limits lead us to very interesting physics.
To understand the physical meaning of the first limit
notice that, with (\ref{fadd}) the Skyrme Lagrangian
reduces to the Skyrme-Faddeev Lagrangian
\bea
{\cal L} \rightarrow -\dfrac{\kappa^2}{2} (\pd_\mu \hn)^2
-\dfrac{\alpha}{4}(\pd_\mu \hn \times \pd_\nu \hn)^2,
\label{sflag}
\eea
whose equation of motion is given by (\ref{sfeq}).
This tells that the Skyrme-Faddeev theory becomes
a self-consistent truncation of the Skyrme theory.
This assures that the Skyrme-Faddeev theory is
an essential ingredient (the backbone) of the Skyrme
theory which describes the core dynamics of Skyrme
theory \cite{prl01,plb04,ijmpa08}.

A remarkable feature of the Skyrme-Faddeev
theory is that it can be viewed as a theory of
monopole. In fact, (\ref{sfeq}) has the singular 
monopole solution \cite{prl01,plb04,ijmpa08}
\bea
\hn = \pm \hat r.
\label{mono}
\eea
which carries the magnetic charge
\bea
Q_m = \dfrac{\pm 1}{8\pi} \int \epsilon_{ijk} 
\big[\hat r\cdot (\pd_i \hat r 
\times \pd_j \hat r)\big] d\sigma_k= \pm 1,
\eea
which represents the homotopy $\pi_2(S^2)$ defined
by $\hn$. Of course, (\ref{mono}) has a point singularity 
at the origin which makes the energy divergent. But 
we can easily regularize the singularity with a non-trivial 
$\om$, with the boundary condition (\ref{skbc}). And 
the regularized monopole becomes nothing but 
the well known skyrmion. 

Moreover, it has the (helical) magnetic vortex solution
made of the monopole-antimonopole pair infinitely
separated apart, and the knot solution which can be
viewed as the twisted magnetic vortex ring made of
the helical vortex whose periodic ends are connected
together. So the monopole plays an essential role in
all these solutions. This shows that the Skyrme-Faddeev
theory, and by implication the Skyrme theory itself,
can be viewed as a theory of monopole.

But perhaps a most important point of the Skyrme-Faddeev
Lagrangian is that it provides a ``missing" link between
Skyrme theory and QCD, because the Skyrme-Faddeev
Lagrangian can actually be derived from
QCD \cite{prl01,plb04,ijmpa08}.

\section{Skyrme Theory and QCD: A Theory of Monopole}

To reveal the deep connection between  Skyrme theory
and QCD, consider the SU(2) QCD
\bea
{\cal L}_{QCD}=-\dfrac14 \vec F_\mn^2.
\label{qcd}
\eea
Now, we can make the Abelian projection choosing the Abelian
direction to be $\hn$ and imposing the Abelian isometry
\bea
D_\mu \hn=0,
\eea
to the gauge potential. With this we obtain the restricted
potential $\hat A_\mu$ which describes the color neutral
binding gluon (the neuron) \cite{prd80,prl81}
\begin{gather}
\A_\mu\rightarrow \hat A_\mu={\cal A}_\mu
+{\cal C}_\mu,   \nn\\
{\cal A}_\mu=A_\mu \hn,
~~~{\cal C}_\mu=-\dfrac1g \hn \times \pd_\mu \hn.
\label{ap}
\end{gather}
The restricted potential is made of two parts, the naive
Abelian (Maxwellian) part ${\cal A}_\mu$ and the topological
monopole (Diracian) part ${\cal C}_\mu$. Moreover, it has
the full non-Abelian gauge freedom although the holonomy
group of the restricted potential is Abelian.

So we can construct the restricted QCD (RCD) made of
the restricted potential which has the full SU(2) gauge
freedom \cite{prd80,prl81}
\begin{gather}
{\cal L}_{RCD} = -\dfrac{1}{4} \hat F^2_\mn  
=-\dfrac{1}{4} F_\mn^2 \nn\\
+\dfrac1{2g} F_\mn \hn 
\cdot (\pro_\mu \hn \times \pro_\nu \hn)  
-\dfrac1{4g^2} (\pro_\mu \hn \times \pro_\nu \hn)^2,  \nn\\
=-\dfrac{1}{4} (F_\mn +H_\mn)^2,  \nn\\
F_\mn=\pd_\mu A_\nu-\pd_\nu A_\mu,  \nn\\
H_\mn= -\dfrac1g \hn \cdot (\pd_\mu \hn \times \pd_\nu \hn)
= \pd_\mu C_\nu - \pd_\nu C_\mu.
\label{rcd}
\end{gather}
This shows that RCD is a dual gauge theory made of 
two Abelian gauge potentials, the electric $A_\mu$
and magnetic $C_\mu$. Nevertheless it has the full 
non-Abelian gauge symmetry.  

Moreover, we can recover the full SU(2) QCD with 
the Abelian decomposition which decomposes 
the gluons to the color neutral neuron and colored 
chromon gauge independently \cite{prd80,prl81}
\begin{gather}
\A_\mu =\hat A_\mu +\vec X_\mu,
~~~\hn \cdot \vec X_\mu=0,   \nn\\
\vec F_\mn=\hat F_\mn+ \hD _\mu \X_\nu 
- \hD_\nu \X_\mu + g\X_\mu \times \X_\nu,
\label{adec}
\end{gather}
where $\vec X_\mu$ describes the gauge covariant 
colored chromon. With this we have the Abelian 
decomposition of QCD 
\begin{gather}
{\cal L}_{QCD} = -\dfrac{1}{4} \vec F^2_\mn
=-\dfrac{1}{4}\hat F_\mn^2-\dfrac{1}{4}(\hD_\mu\X_\nu
-\hD_\nu\X_\mu)^2 \nn\\
-\dfrac{g}{2} {\hat F}_{\mu\nu} \cdot (\X_\mu \times \X_\nu)
-\dfrac{g^2}{4} (\X_\mu \times \X_\nu)^2. 
\label{ecd} 
\end{gather}
This confirms that QCD can be viewed as RCD made of 
the binding gluon, which has the colored valence gluon 
as its source \cite{prd80,prl81}. The Abelian decomposition
has been known as the Cho decomposition, Cho-Duan-Ge
(CDG) decomposition, or Cho-Faddeev-Niemi (CFN) 
decomposition in the literature \cite{fadd,gies,kondo,zucc}

Now we can reveal the connection between the Skyrme 
theory and QCD. Let us start from RCD and assume that
the binding gluon (restricted potential) acquires a mass
term after the confinement. In this case (\ref{rcd}) becomes
\bea
&{\cal L}_{RCD}\rightarrow  -\dfrac{1}{4} \hat F^2_\mn
-\dfrac{m^2}{2} \hat A_\mu^2  \nn\\
&=-\dfrac{1}{4} F_\mn^2
+\dfrac1{2g} F_\mn \hn \cdot (\pd_\mu \hn \times \pd_\nu \hn)  \nn\\
&-\dfrac1{4g^2} (\pd_\mu \hn \times \pd_\nu \hn)^2
-\dfrac{m^2}{2} \big[A_\mu^2+ \dfrac1{g^2} (\pd_\mu \hn)^2\big].
\label{mrcd}
\eea
Of course, the mass term breaks the gauge symmetry, 
but this can be justified because the binding gluons 
could acquire mass after the confinement sets in.

Integrating out the $A_\mu$ potential (or simply putting
$A_\mu=0$), we can reduce (\ref{mrcd}) to
\bea
&{\cal L}_{RCD}\rightarrow-\dfrac1{4g^2}
(\pd_\mu \hn \times \pd_\nu \hn)^2
-\dfrac{m^2}{2g^2} (\pd_\mu \hn)^2,
\label{mrcd1}
\eea
which becomes nothing but the Skyrme-Faddeev 
Lagrangian (\ref{sflag}) when $\kappa^2=m^2/g^2$ 
and $\alpha=1/g^2$. This tells that the Skyrme-Faddeev 
Lagrangian can actually be derived from RCD. And this 
is a mathematical derivation. This confirms that the Skyrme
theory and QCD is closely related, more closely than it
appears.

This has deep consequences. Notice that $\hn$ which
provides the Abelian projection in QCD naturally
represents the monopole topology $\pi_2(S^2)$, and
can describe the monopole. In fact, it is well known
that $\hn=\hat r$ becomes exactly the Wu-Yang
monopole solution \cite{prl80}. On the other hand 
the above exercise tells that this $\hn$ is nothing 
but the normalized pion field in the Skyrme theory. 
This strongly implies that $\hn=\hat r$ can also 
describe the monopole solution in the Skyrme theory.  
Indeed, (\ref{mono}) is precisely the Wu-Yang
monopole transplanted in the Skyrme theory.

What is more, the Skyrme theory has more complicated
monopole solutions. With $\hn=\hat r$ and the boundary
condition
\bea
\omega(0)=\pi,~~~~\omega(\infty)=0,~~~~({\rm mod}~2\pi)
\label{+mbc}
\eea
we can find a monopole solution which reduces to the singular
solution (\ref{mono}) near the origin. The only difference
between this and (\ref{mono}) is the non-trivial dressing of
the scalar field $\omega$, so that it could be interpreted as
a dressed monopole. This dressing, however, is only partial
because this makes the energy finite at the infinity, but not
at the origin.

Similarly, with the following boundary condition
\bea
\omega(0)=2\pi,~~~~\omega(\infty)=\pi,~~~~({\rm mod}~2\pi)
\label{-mbc}
\eea
we can obtain another monopole and half-skyrmion solution
which approaches the singular solution near the infinity.
Here again the partial dressing makes the energy finite at
the origin, but not at the infinity. So the partially dressed
monopoles still carry an infinite energy.

\begin{figure}
\begin{center}
\includegraphics[height=4cm, width=7cm]{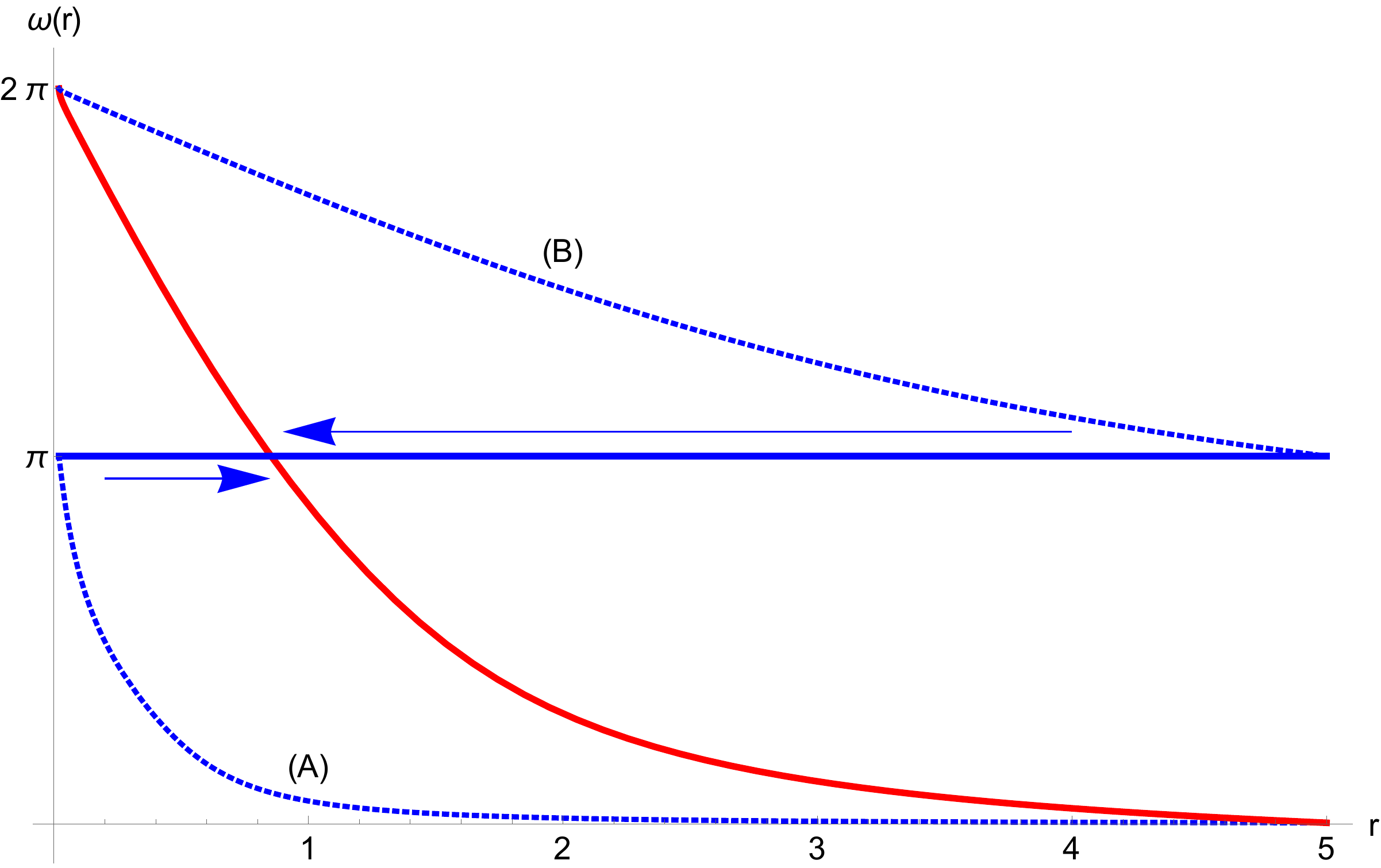}
\caption{\label{mosky} The singular Wu-Yang monopole 
(the blue line), the two half skyrmions (A) and (B) as 
partially dressed singular monopoles (blue curves), 
and the regular skyrmion (red curve) obtained combining 
the two half skyrmions.}
\end{center}
\end{figure}

Clearly these solutions carry the unit magnetic charge of
the homotopy $\pi_2(S^2)$ defined by $\hn$ \cite{ijmpa08}
\bea
&M = \dfrac{\pm 1}{8\pi} \int \epsilon_{ijk} \big[\hat r
\cdot (\pd_i \hat r \times \pd_j \hat r)\big] d\sigma_k
= \pm 1,
\label{mone}
\eea
but carry a half baryon number
\bea
&B= -\dfrac{1}{8\pi^2} \int \epsilon_{ijk} \pd_i \om
\big[\hat r \cdot (\pd_j \hat r \times \pd_k \hat r) \big]
\sin^2 \dfrac{\om}{2} d^3r \nn\\
&=-\dfrac{1}{\pi} \int \sin^2 \dfrac{\om}{2} d\om=\dfrac12.
\label{bnhalf}
\eea
This is due to the boundary conditions (\ref{+mbc}) and
(\ref{-mbc}). So it describes a half-skyrmion.

The half-skyrmions have two remarkable features.
First, combining the two half-skyrmions we can
form a finite energy soliton, the fully dressed
skyrmion \cite{ijmpa08}. This must be clear because,
putting the boundary conditions (\ref{+mbc})
and (\ref{-mbc}) together we recover the boundary
condition (\ref{skbc}) of the skyrmion. So putting
the two partially dressed monopoles together we
obtain the well known finite energy skyrmion solution.

This is summarized in Fig. \ref{mosky}, where the blue line
represents the singular monopole, the dotted curves (A) and
(B) represent the singular half skyrmions, and the red curve
represents the regular skyrmion. Notice that in all these
solutions we have $\hn=\hr$, so that $M=1$. This confirms
that the skyrmion is nothing but the Wu-Yang monopole
of QCD, transplanted and regularized to have a finite
energy by the massless scalar field $\om$ in the Skyrme
theory \cite{prl01,plb04,ijmpa08}.

The other remarkable feature of the half-skyrmions
is that the monopole number (\ref{mone}) and
the baryon number (\ref{bnhalf}) are different.
This is very interesting because, in skyrmions
the baryon number is commonly identified by
the rational map which defines the monopole
number. This suggests that the baryon number 
and the monopole number are one and the same
thing \cite{man}. But obviously this is not true
for the half-skyrmions. In the following we will 
argue that this need not be ture, and show that 
the skyrmions in general have two different topological 
numbers.

\begin{figure}
\begin{center}
\includegraphics[width=6cm, height=4cm]{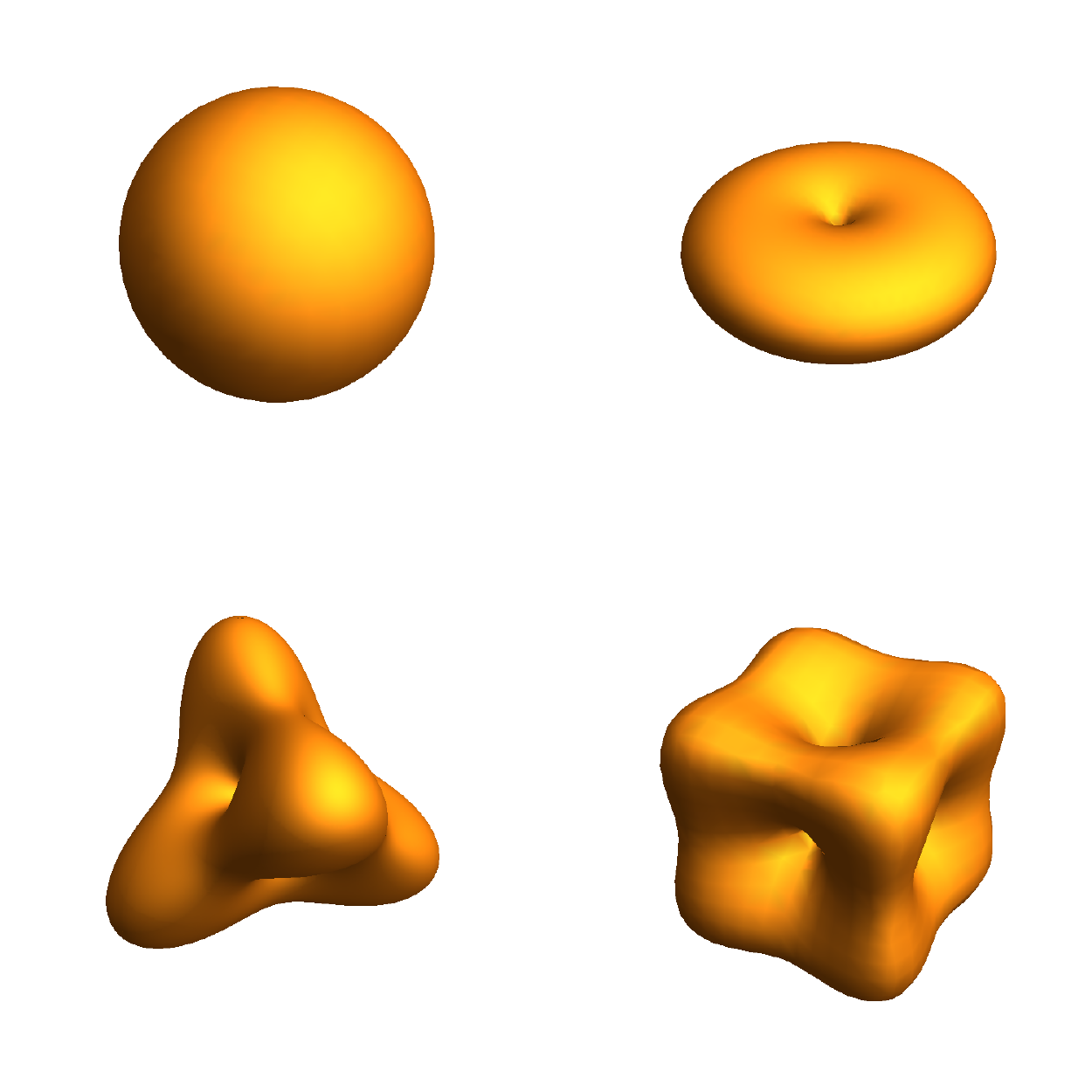}
\caption{\label{oldsoln} The well known (non spherically
symmetric) numerical multi-skyrmion solutions with
baryon number 1,2,3,4.}
\end{center}
\end{figure}

\section{Skyrmions with Two Topological Numbers}

The above argument clearly shows that the skyrmion has
the monopole topology. And this monopole topology
reappear in all popular non spherically symmetric
multi-skyrmion solutions \cite{man,bat,sut}.
Indeed, it is well known that the monopole topology of
the rational map $\pi_2(S^2)$ given by $\hn$,
\bea
&M= \dfrac{1}{8\pi} \int \epsilon_{ijk} \big[\hn
\cdot (\pd_i \hn \times \pd_j \hn)\big] d\sigma_k
= m,
\label{mnm}
\eea
together with the boundary condition (\ref{skbc}),
has played the fundamental role for many people
to construct these multi-skyrmion solutions. Some
of these multi-skyrmion solutions are shown in 
Fig. \ref{oldsoln}.

On the other hand, these solutions have always been
interpreted to have the baryon topology $\pi_3(S^3)$
which have the baryon number given by
\bea
&B= -\dfrac{1}{8\pi^2} \int \epsilon_{ijk} \pd_i \om
\big[\hn \cdot (\pd_j \hn \times \pd_k \hn) \big]
\sin^2 \dfrac{\om}{2} d^3r \nn\\
&= \dfrac{1}{8\pi} \int \epsilon_{ijk} \big[\hn
\cdot (\pd_i \hn \times \pd_j \hn)\big] d\sigma_k
=m.
\label{bnm}
\eea
But obviously this baryon number is precisely the monopole
number shown in (\ref{mnm}), so that these solutions
have $B=M$. Nevertheless, the monopole topology
$\pi_2(S^2)$ and the baryon topology $\pi_3(S^3)$ is
clearly different. If so, we may ask what is the relation
between the monopole topology and the baryon topology.

\begin{figure}
\begin{center}
\includegraphics[height=4cm, width=7cm]{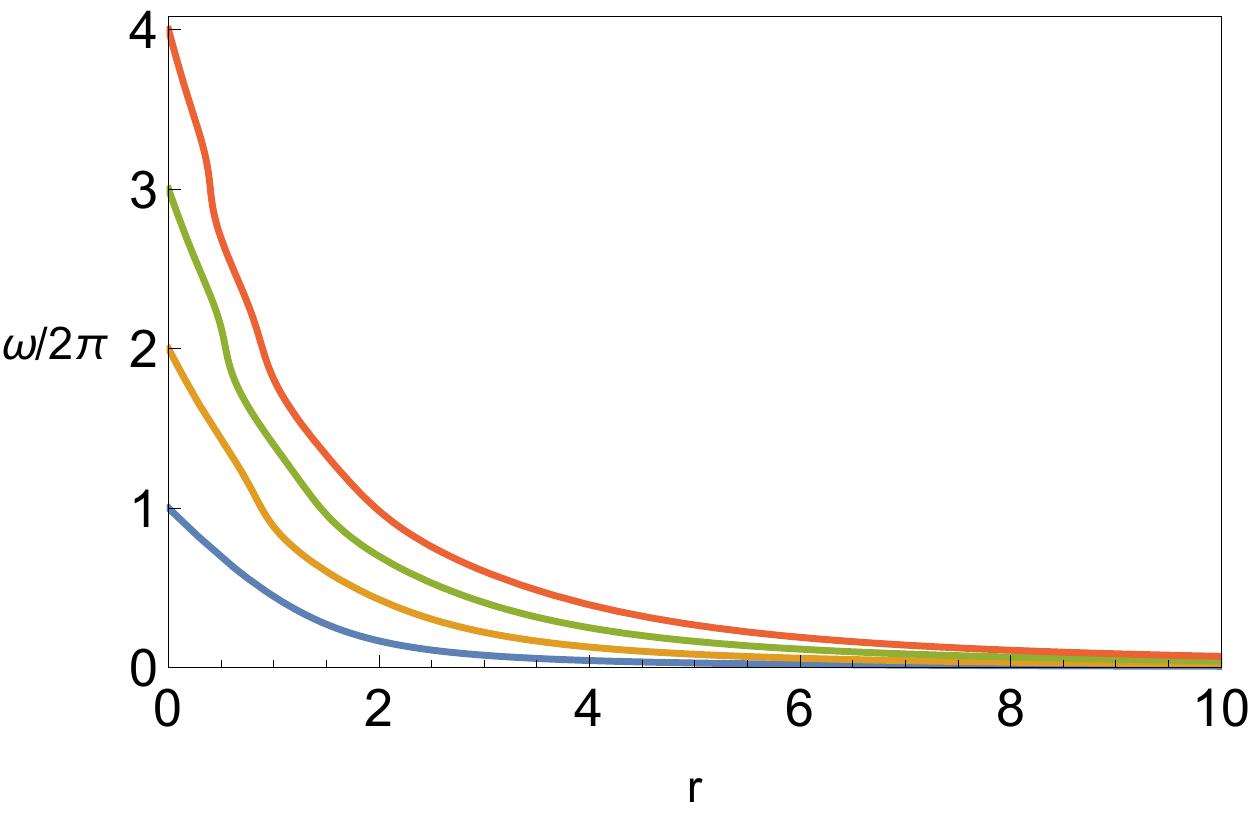}
\caption{\label{skysol} The spherically symmetric skyrmions
with baryon number 1,2,3,4, which should be contrasted
with the solutions shown in Fig. \ref{oldsoln}.}
\end{center}
\end{figure}

The Skyrme equation (\ref{skeq2}) in the spherically
symmetric limit plays an important role to clarify this
point. To see this notice that, although the SU(2)
matrix $U$ is periodic in $\om$ variable by $4\pi$,
$\om$ itself can take any value from $-\infty$ to
$+\infty$. So we can obtain the spherically symmetric
multi-skyrmion solutions generalizing the boundary
condition (\ref{skbc}) to \cite{sky,witt,prep}
\bea
\omega(0)= 2\pi n,~~~~~\omega(\infty)= 0,
\label{skbcn}
\eea
with an arbitrary integer $n$. Some of the spherically
symmetric multi-skyrmion solutions are shown in
Fig. \ref{skysol}. These are, of course, the solutions
that Skyrme originally proposed to identify as the nuclei
with baryon number larger than one \cite{sky}. But
soon after they are dismissed as uninteresting because
they have too much energy.

An interesting features of the spherically symmetric
solutions is that whenever the curve passes through
the values $\om=2\pi n$, it become a bit steeper.
This is because (as we will see soon) these points are
the vacua of the theory, and the steep slopes shows
that the energy likes to be concentrated around
these vacua.

\begin{figure}
\begin{center}
\includegraphics[height=4cm, width=7cm]{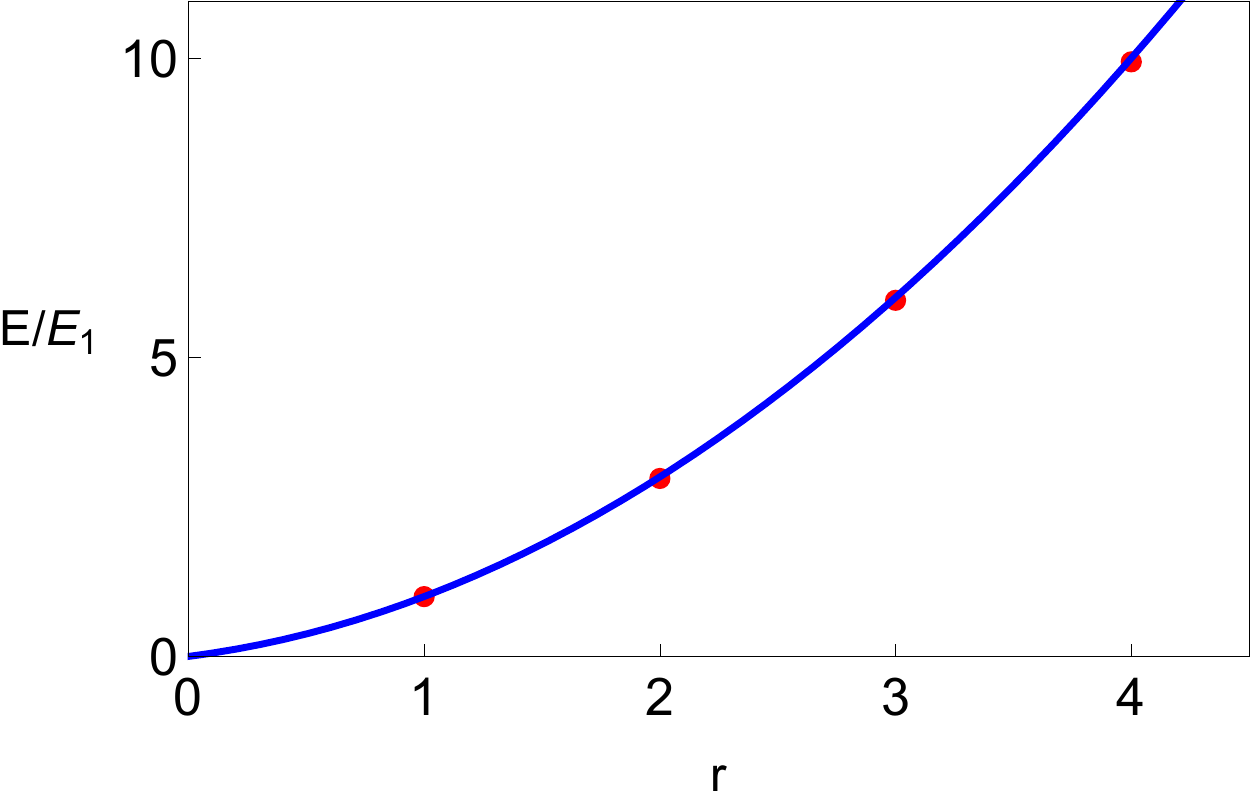}
\caption{\label{En} The energy of the spherically symmetric
solutions with baryon number 1,2,3,4. The numerical
fit (the blue curve) and the $n(n+1) E_1/2$ curve (the green
curve) are almost indistinguishable.}
\end{center}
\end{figure}

Another interesting feature of the spherically symmetric
skyrmions is the energy, which is given by \cite{ijmpa08}
\bea
&E =\dfrac{\pi \kappa^2}{2}  \Int_0^\infty \Big\{\Big(r^2
+\dfrac{2\alpha}{\kappa^2} \sin^2\dfrac{\om}{2}\Big)
\Big(\dfrac{d\om}{dr}\Big)^2  \nn\\
&+8 \Big(1+\dfrac{\alpha}{2\kappa^2 r^2} \sin^2
\dfrac{\om}{2} \Big) \sin^2 \dfrac{\om}{2} \Big\} dr  \nn\\
&= \pi {\sqrt \alpha} \kappa \Int^{\infty}_{0}
\Big[x^2 \left(\dfrac{d\om}{dx}\right)^2
+ 8 \sin^2{\dfrac{\om}{2}} \Big] dx,
\label{sken}
\eea
where $x=\kappa r/{\sqrt \alpha}$. From this we can easily
calculate their energy. This is shown in Fig. \ref{En}.

Numerically the baryon number dependence of the energy
is given by \cite{sky,prep}
\bea
E_n\simeq \dfrac{n(n+1)}{2} E_1.
\label{en}
\eea
This has two remarkable points. First, the baryon number
dependence of the energy is quadratic. This, of course,
means that the energy of the skyrmion with baryon number
$n$ is bigger than the sum of $n$ lowest energy skyrmion.
So the radially excited skyrmions are unstable. This is
not the case in the popular multi-skyrmion solutions shown
in Fig. \ref{oldsoln}. They have positive binding energy,
so that the energy of the skyrmion with baryon number $B$
becomes smaller than the $B$ sum of the lowest energy
skyrmion. This was the main reason why the spherically
symmetric solutions have not been considered seriously
as the model of heavy nuclei.

But actually (\ref{en}) makes the spherically symmetric
solutions mathematically more interesting, because
(\ref{en}) is almost exact. In fact Fig. \ref{En} shows
that the mathematical curve $E_n= n(n+1)E_1/2$ and
the numerical fit is almost indistinguishable. We could 
understand this as follows. Roughly speaking, the kinetic 
energy (first part) and the potential energy (second 
part) of (\ref{sken}) become proportional to $n^2$ and 
$n$, and the two terms have an equal contribution due 
to the equipartition of energy. But we need a better 
explanation of (\ref{en}).

\begin{figure}
\begin{center}
\includegraphics[height=4cm, width=7cm]{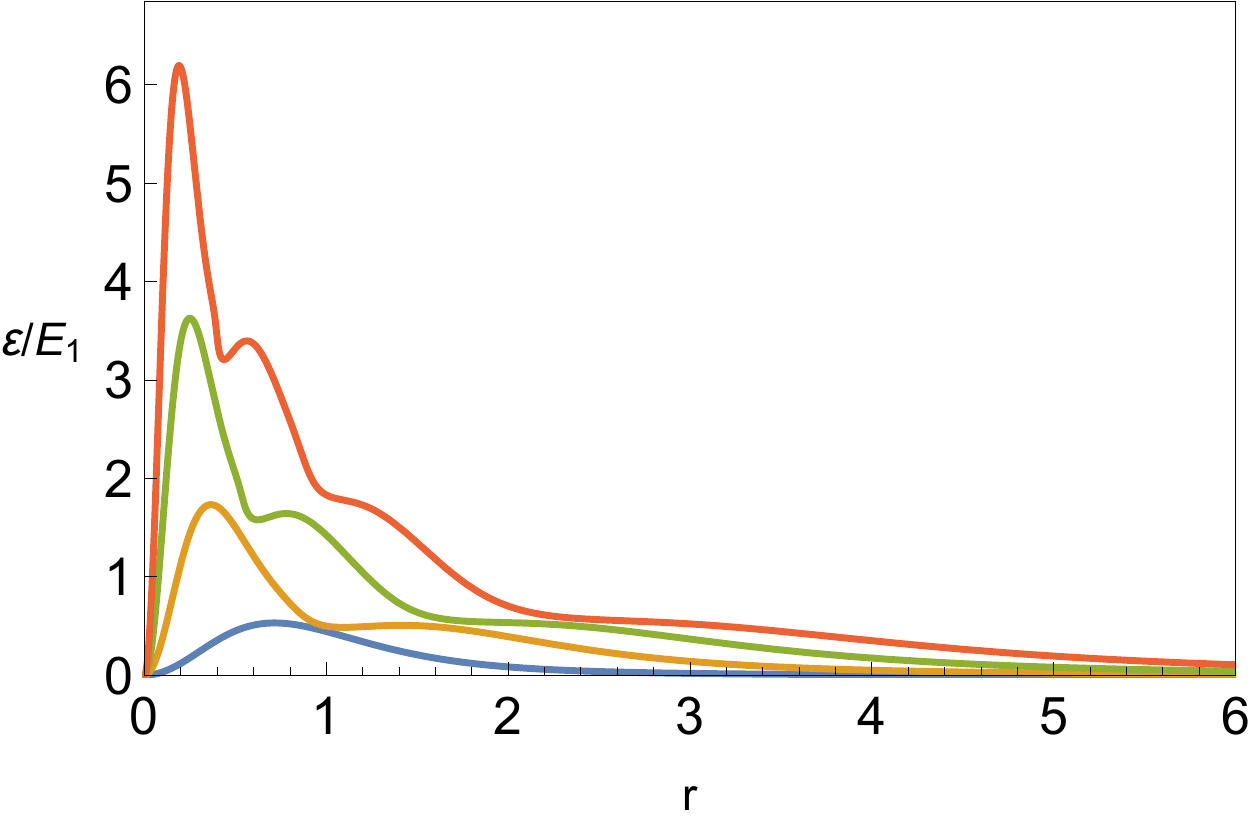}
\caption{\label{eden} The radial energy density of
the spherically symmetric skyrmions with baryon number
1,2,3,4. The density functions are normalized
to make the integral of the $B=1$ solution to be
the unit.}
\end{center}
\end{figure}

Fig. \ref{eden} shows the energy density of the solutions. 
Clearly the $B=n$ solution has $n$ local maxima, which 
indicates that it is made of n shells of unit skyrmions. 
Moreover, as we have remarked the energy density has 
the local maxima at $\om=2\pi n$. So they describe the shell 
model of nuclei, made of $n$ shells located at $\om=2\pi n$. 
This tells that the spherically symmetric solutions could 
be viewed as radially extended skyrmions of the original 
skyrmion. In this interpretation the baryon number $n$ 
of these skyrmions can be identified as the radial, or 
more properly the shell, number \cite{piet}. And this
baryon number is fixed by the winding number $\pi_1(S^1)$ 
of the angular variable $\om$.

The contrast between the popular non spherically
symmetric solutions shown in Fi‍g. \ref{oldsoln}
and the sphrically symmetric solutions shown in
Fig. \ref{skysol} is unmistakable. But the contrast
is not just in the appearance. They are fundamentally
different. In particular, the spherically symmetric
solutions have a very important implication.

To understand this, notice that these solutions have
the monopole number given by
\bea
&M= \dfrac{1}{8\pi} \int \epsilon_{ijk} \big[\hr
\cdot (\pd_i \hr \times \pd_j \hr)\big] d\sigma_k =1.
\label{mn1}
\eea
On the other hand, their baryon number is given by 
the winding number $\pi_1(S^1)$ of $\om$ determined 
by the boundary condition (\ref{skbcn}),
\bea
&B= -\dfrac{1}{8\pi^2} \int \epsilon_{ijk} \pd_i \om
\big[\hat r \cdot (\pd_j \hat r \times \pd_k \hat r) \big]
\sin^2 \dfrac{\om}{2} d^3r \nn\\
&=-\dfrac{1}{\pi} \int \sin^2 \dfrac{\om}{2} d\om=n.
\label{bnn}
\eea
So, unlike the popular non spherically symmetric
multi-skyrmions shown in Fig. \ref{oldsoln}, the baryon
number and the monopole number of these solutions
are different. Moreover, the baryon number is not given
by the rational map, but by the winding number.

This confirms that the skyrmions actually do carry two
topological numbers, the baryon number of $\pi_3(S^3)$
and the monopole number of $\pi_2(S^2)$ \cite{epjc17}.
And the spherically symmetric solutions play the crucial
role to demonstrate this.

In this scheme the skyrmions are classified by
$(m,b)$, the monopole number $m$ and the baryon 
number $b$. And the popular (non spherically 
symmetric) solutions become the $(m,m)$ skyrmions 
and the spherically symmetric solutions become 
the $(1,n)$ skyrmions.

This has a deep consequence. Based on this we
can actually show that the baryon number can be
decomposed to the monopole number and the radial
(shell) number and expressed by the product of
the two numbers \cite{epjc17}. To show this we
discuss the characteristic features of the spherically
symmetric solutions first.

This implies that we could replace the baryon number
with the shell number, and classify the skyrmions by
the monopole number $m$ and shell number $n$ by
$(m,n)$, instead of $(m,b)$. In other words we could
also classify the skyrmions by the monopole topology
$\pi_2(S^2)$ of $\hn$ and the U(1) (i.e., shell) topology
$\pi_1(S^1)$ of $\om$. In this scheme the baryon number
of the $(m,n)$ skyrmion is given by $B=mn$. This must be
clear from (\ref{bn}), which tells that the baryon number
is made of two parts, the $\pi_2(S^2)$ of $\hn$ and
$\pi_1(S^1)$ of $\om$.

According to this classification the popular (non spherically
symmetric) skyrmions become the $(m,1)$ skyrmions,
the radially extended spherically symmetric skyrmions
become the $(1,n)$ skyrmions. The justification for this
classification comes from the following observation. First,
the $S^3$ space (both the real space and the target space)
in the baryon topology $\pi_3(S^3)$ admits the Hopf
fibering $S^3\simeq S^2\times S^1$. Second, the two
variables $\hn$ and $\om$ of the Skyrme theory naturally
accommodate the monopole topology $\pi_2(S^2)$ and
the shell topology $\pi_1(S^1)$.

Furthermore, in this classification the baryon number
is given by the product of the monopole number and
the shell number, or $B=mn$. To see this notice that
the Hopf fibering of $S^3$ has an important property
that locally it can be viewed as a Cartesian product of 
$S^2$ and $S^1$. So we have
\bea
&B =-\dfrac{1}{8\pi^2} \Int \epsilon_{ijk} \pd_i \om
\big[\hn \cdot (\pd_j \hn \times \pd_k \hn) \big]
\sin^2 \dfrac{\om}{2} d^3r \nn\\
&=-\dfrac{1}{8\pi^2} \Int \pd_i \om
\big[\hn \cdot (\pd_j \hn \times \pd_k \hn) \big]
\sin^2 \dfrac{\om}{2} dx^i \wedge dx^j \wedge dx^k \nn\\
&=\dfrac{n}{8\pi} \Int \epsilon_{ijk} \big[\hn \cdot
(\pd_i \hn \times \pd_j \hn) \big] d\Sigma_k=mn,
\label{bmn}
\eea
where $d\Sigma_k=\epsilon_{ijk} dx^i \wedge dx^j/2$.
This shows that the baryon number of the skyrmion
is given by the product of the monopole number and
the shell number. Obviously both $(m,1)$ and $(1,n)$
skyrmions are the particular examples of this.

Another way to show this is to introduce a generalized
coordinates $(\eta,\alpha,\beta)$ for $R^3$ whose line
element is given by 
\begin{gather}
ds^2= f(\eta) d \eta^2 +d \Sigma^2(\alpha,\beta),
\end{gather}
in which $\eta$ represents the radial coordinate $R^1$
and $(\alpha,\beta)$ represents the $S^2$ such that
$\om=\om(\eta)$ and $\hn=\hn(\alpha,\beta)$. Again this
is possible because the hopf fibering of $S^3$ can be
expressed by a locally Cartesian product of $S^2$ and
$S^1$. In this coordinates we clearly have
\begin{gather}
B =-\dfrac{1}{8\pi^2} \Int \epsilon_{ijk} \pd_i \om
\big[\hn \cdot (\pd_j \hn \times \pd_k \hn) \big]
\sin^2 \dfrac{\om}{2} d^3x \nn\\
=-\dfrac{1}{4\pi^2} \Int \sin^2 \frac{\om}{2} d \om
\Int \big[\hn \cdot (d \hn \wedge d\hn) \big] 
=mn.
\label{bnmn}
\end{gather}
So in this local direct product coordinates, it is
straightforward to prove $B=mn$. Notice that
the third equality tells that the baryon number
can be expressed in a coodinate independent form.

Clearly the spherically symmetric solutions satisfy
this criterion. In this case, $\pi_3(S^3)$ naturally
decomposes to $\pi_2(S^2)$ of $\hn$ and $\pi_1(S^1)$
of $\om$, so that the baryon number $B$ is decomposed
to the $S^2$ wrapping number $m$ and the $S^1$
winding number $n$. But the spherical symmetry
requires $\hn=\hat r$, and restricts $m=1$.

Again this follows from two facts. First, the fact that
the Hopf fibering allows $S^3$ to decompose $S^2 \times S^1$,
and the fact that in Skurme theory $\om$ and $\hn$ naturally
decompose the target space $S^3$ to $S^1$ and $S^2$. With
this we have $\pi_3(S^3) \simeq \pi_2(S^2) \times \pi_1(S^1)$.
This strongly support that the baryon topology $\pi_3(S^3)$
can be refined to $\pi_2(S^2)$ of $\hn$ and $\pi_1(S^1)$
of $\om$.

If so, one might wonder whether the popular skyrmions
can also be made to carry two different topological
numbers. This should be possible. To see this remember
that the integer $n$ in the $(1,n)$ skyrmions describes
the radial (shell) number which describes the shell
structure of the spherically symmetric skyrmions. And
this shell structure is provided by the angular variable
$\om$ which allows the radial extension of the original
$(1,1)$ skyrmion. So we can generalize the $(m,1)$
skyrmion to have similar shell structure.

In fact, we may obtain the ``radially extended" solutions
of the non spherically symmetric skyrmions numerically,
generalizing the boundary condition (\ref{skbc}) to
(\ref{skbcn}), requiring \cite{piet}
\bea
&\om(r_k)=2\pi k,~~~(k=0,1,2,...n),  \nn\\
&r_0=0 ~\langle~r_1~\langle~...~\langle~r_n=\infty,
\eea
keeping the rational map number $m$ of $\hn$ unchanged.
With this we could find new solutions numerically minimizing
the energy, varying $r_k~(k=1,2,...,n-1)$. This way we can
add the shell structure and the shell number to the $(m,m)$
skyrmion. In this case the baryon number of the radially
extended skyrmions should become $B=mn$.

But the above argument tells that in general (even for non
spherically symmetric skyrmions) we may introduce the
generalized coordinates such that $\hn=\hn(\alpha,\beta)$
and $\om=\om(\gamma)$, and can still make the ``radial"
extension of the skyrmions to obtain the radially extended
$(m,n)$ skyrmions which have $n$ shells numerically,
generalizing the boundary condition (\ref{skbc}) to
(\ref{skbcn}).

Again this becomes possible because the Skyrme theory is
described by two variables $\hn$ and $\om$ which naturally
accommodate the monopole topology $\pi_2(S^2)$ and
the shell topology $\pi_1(S^1)$.  This is very important,
because mathematically there is no way to justify the replacement
of the $\pi_3(S^3)$ topology by two independent $\pi_2(S^2)$
topology and $\pi_1(S^1)$ topology.

Now, one may ask about the stability of the $(m,n)$ skyrmions.
Clearly an $(m,n)$ skyrmion does not have to be stable, and
could decay to lower energy skyrmions as far as the decay
is energetically allowed. For example, (\ref{en}) clearly
tells that the $(1,n)$ skyrmion can decay to lower energy
skyrmions. This is natural. An interesting and important
question here is whether the two topological numbers $m$
and $n$ are conserved independently or not. Certainly
the baryon topology and the monopole topology are mathematically
independent. This means that  the baryon number and
the monopole number must be conserved separately. This
(with $\delta B=m~\delta n+ n~\delta m$) automatically
guarantees that the shell number must also be conserved.

This means that the two numbers $m$ and $n$ must be
conserved separately, so that the $(n,m)$ skyrmion could
decay to $(n_1,m_1)$ and $(n_2,m_2)$ skyrmions when
$n=n_1+n_2$ and $m=m_1+m_2$. But there is no way that
the two topological numbers $m$ and $n$ can be transformed
to each other. This tells that the skyrmions retains
the topological stability of two topology independently,
even when they are classified by two topologial numbers.

The above discussions raise another deep question.
As we have remarked, when $\om=(2n+1)\pi$, 
the Skyrme theory reduces to the Skyrme-Faddeev 
theory. In this limit the Skyrme theory has the knot 
solutions described by $\hn$ whose topology is given 
by $\pi_3(S^2)$ \cite{prl01,plb04,ijmpa08}. And in 
these knots, $\om$ is not activated. If so, one might 
ask if we can dress the knots with $\om$ and add 
the shell structure to the knots to have two topological 
numbers $\pi_1(S^1)$ and $\pi_3(S^2)$. This is a mind 
boggling question which certainly deserves more study.

\section{Multiple Vacua of Skyrme Theory}

Skyrme theory has been known to have rich topological
structures. It has the Wu-Yang type monopoles which have
the $\pi_2(S^2)$ topology, the skyrmions which have
the $\pi_3(S^3)$ topology, the baby skyrmions which have
the $\pi_1(S^1)$ topology, and the Faddeev-Niemi knots which
have the $\pi_3(S^2)$ topology \cite{prl01,plb04,ijmpa08}.
In the above we have shown that the theory has more
topological structure, and proved that the skyrmions
can be generalized to have two topological quantum
numbers. But this is not the end of story.

The Skyrme theory in fact has another very important 
topological structure, the topologically different 
multiple vacua. To see this, notice that
(\ref{skeq1}) has the solution
\bea
\om=2\pi p,~~~(p;~integer),
\label{skyvac}
\eea
independent of $\hn$. Clearly this is the vacuum
solution.

This shows that the Skyrme theory has multiple vacua
classified by the integer $p$. This is similar to the vacuum 
of the Sine-Gordon theory, but we emphasize that, unlike 
the Sine-Gordon theory, the Skyrme theory has the multiple 
vacua without any potential. Of course, we could have such 
vacua in Skyrme theory introducing a potential term in 
the Lagrangian. This is not what we are doing here. Another 
pont is that the spherically symmetric skyrmions occupy 
and connect the $p+1$ adjacent vacua. This tells that 
we can connect all vacua with the spherically symmetric 
skyrmions.

Perhaps more importantly, (\ref{skyvac}) becomes 
the vacuum independent of $\hn$, which is completely 
arbitrary. So $\hn$ can add the $\pi_3(S^2)$ topology 
to each of the multiple vacua classified by another 
integer $q$. And this topology is precisely the knot 
topology of the QCD vacuum \cite{plb07}.

This is not surprising. In view of the deep connection 
between QCD and Skyrme theory, it is natural that
they have similar vacuum structure. To clarify this 
point we introduce a right-handed unit isotriplet 
$(\hn_1,\hn_2,\hn_3=\hn)$, and impose the vacuum
isometry to the SU(2) gauge potential,
\bea
^{\forall_i} D_\mu \hn_i=0,
\eea
which assures $\vec F_\mn=0$. From this we obtain
the most general SU(2) QCD vacuum potential \cite{plb07}
\bea
&\A_\mu\rightarrow \hat \Omega_\mu
=-\dfrac12 \epsilon_{ijk} (\hn_i\cdot \pd_\mu \hn_j)
~\hn_k  \nn\\
&=\dfrac12 \epsilon_{ijk} (\hn_i\cdot \pd_\mu \hn_j)
~\hat e_k,
\label{qcdvac}
\eea
where $\hat e_1=(1,0,0),\hat e_2=(0,1,0),\hat e_3=(0,0,1)$.

This is the QCD vacuum which has the knot topology
$\pi_3(S^3)\simeq \pi_3(S^2)$. It is clear that
(\ref{qcdvac}) describes the $\pi_3(S^3)$ topology,
since $(\hn_1,\hn_2,\hn)$ defines the mapping
$\pi_3(S^3)$ from the compactified 3-dimensional
space to the SU(2) group space. But notice that
$(\hn_1,\hn_2,\hn)$ is completely determined by
$\hn$, up to the U(1) rotation which leaves $\hn$
invariant, which describes the mapping from the real
space $S^3$ to the coset space $S^2$ of $SU(2)/U(1)$.
So the QCD vacuum (\ref{qcdvac}) can also be classified
by the knot topology $\pi_3(S^2)$ \cite{plb07}.

Now it becomes clear why the Skyrme theory has the same
knot topology. As we have noticed, the Skyrme theory
has the vacuum (\ref{skyvac}), independent of $\hn$.
But this $\hn$ (just as in the SU(2) QCD) defines
the mapping $\pi_3(S^2)$, with the $S^3$ compactification
of $R^3$, and thus can be classified by the knot topology.
Of course, in the Skyrme theory we do not need the
vacuum potential (\ref{qcdvac}) to describe the vacuum.
All we need to describe the knot topology is $\hn$. 

This tells that the vacuum of the Skyrme theory has 
the topology of the Sine-Gordon theory and QCD combined 
together. So the vacuum of the Skyrme theory can also be
classified by two topologcal numbers $(p,q)$, the $\pi_1(S^1)$
of $\om$ and $\pi_3(S^2)$ of $\hn$. And this is so without
any extra potential. As we know, there is no other
theory which has this type of vacuum topology.

The multiple vacua leads us to the question if the
Skyrme theory has a vacuum tunneling which 
connects these vacua. Certainly this is a very interesting 
question worth to be studied further. 

Before we close we emphasize the followings. First, the knot
topology $\pi_3(S^2)$ of $\hn$ of the vacuum is different 
from the monopole topology $\pi_2(S^2)$ of $\hn$. 
The monopole topology is associated to the isolated 
singularities of $\hn$, but for the knot topology $\hn$ 
does not have any singularity for. For the vacuum $\hn$ 
must be completely regular everywhere. 

Second, the knot topology of the vacuum is different
from the Faddeev-Niemi knot in the Skyrme
theory \cite{prl01}. The Faddeev-Niemi knot is a real 
knot which carries energy, but the vacuum knot has 
no energy. Moreover, we have the knot solution when 
$\om=(2n+1)\pi$, but we have the knot of the vacuum 
when $\om= 2\pi p$. In addition, there are infinitely 
many $\hn$ which describes the same vacuum knot 
topology, while the Faddeev-Niemi knot is unique.

What is really remarkable here is that the same $\hn$ has
multiple roles. It describes the monopole topology of 
the skyrmion, the knot topology of Faddeev-Niemi knot, 
and the knot topology of the vacuum.

\section{Knot in Skyrme Theory}

It is well known that in Skyrme theory has the baby skyrmions,
the magnetic vortex made of monopole-antimonopole pair 
infinitely separated apart \cite{piet}. Moreover, with this 
we can construct the helical baby skyrmion, twisting 
the magnetic vortex \cite{ijmpa08}. But the helical vortex 
becomes unstable and decays to the untwisted baby 
skyrmion, unless the periodicity condition is enforced 
by hand. A natural way to make the helical baby skyrmion 
stable is to make it a knot connecting two periodic ends 
and making it a twisted magnetic vortex ring. 

By construction this twisted vortex 
ring carries two magnetic fluxes, $m$ unit of flux 
passing through the disk of the ring and $n$ unit 
of flux moving along the ring. Moreover the two fluxes 
can be thought of two closed rings linked together 
winding each other $m$ and $n$ times, whose linking 
number becomes $mn$. This tells that the twisted 
vortex ring becomes a knot \cite{prl01,plb04,ijmpa08}.

To confirm this we solve the knot equation (\ref{sfeq}) 
explicitly with a consistent ansatz. We first adopt 
the toroidal coordinates $(\eta,\gamma,\varphi)$ given 
by
\bea
&x=\dfrac{a}{D}\sinh{\eta}\cos{\varphi},
~~~y=\dfrac{a}{D}\sinh{\eta}\sin{\varphi}, \nn\\
&z=\dfrac{a}{D}\sin{\gamma},
~~~D=\cosh{\eta}-\cos{\gamma}, \nn\\
&ds^2=\dfrac{a^2}{D^2} \Big(d\eta^2+d\gamma^2+\sinh^2\eta
d\varphi^2 \Big), \nn\\
&d^3x=\dfrac{a^3}{D^3} \sinh{\eta} d\eta d\gamma d\varphi,
\label{tc}
\eea
where $a$ is the radius of the knot defined by $\eta=\infty$. 
With this we choose the following axially symmetric ansatz
\bea
&\hn=\left(\begin{array}{cc} \sin f \cos (n\beta+m\varphi) \\ 
\sin f \sin (n\beta + m\varphi) \cr \cos f \end{array} \right),
\label{skkans}
\eea
where $f$ and $\beta$ are functions of $\eta$ and $\gamma$. 

With this we have
\bea
&C_\mu = n(\cos f -1) \pd_\mu \beta 
+m(\cos f +1)\pd_\mu \varphi, \nn\\
&H_{\eta \gamma}=-nK \sin f,
~~~~H_{\gamma \varphi }=-m\sin f\pd_\gamma f, \nn\\
&H_{\varphi \eta }=m \sin f\pd_\eta f, \nn\\
&K = \pd_\eta f\pd_\gamma \beta -\pd_\gamma f \pd_\eta \beta,
\label{kmf1}
\eea
so that the knot equation (\ref{sfeq}) is written as
\bea
&\Big[\partial_\eta^2 +\pd_\gamma^2
+\Big(\dfrac{\cosh \eta}{\sinh \eta}
-\dfrac{\sinh \eta}D\Big)\pd_\eta
-\dfrac{\sin \gamma}D\pd_\gamma \Big]f  \nn\\
&-\Big(n^2 \big((\pd_\eta \om)^2
+(\pd_\gamma \om)^2 \big)
+\dfrac{m^2}{\sinh ^2\eta}\Big) \sin f\cos f  \nn \\
&+\dfrac{\alpha}{\kappa^2} \dfrac{D^2}{a^2}
\Big(A\cos f +B\sin f \Big)\sin f =0, \nn\\
&\Big[\pd_\eta^2 +\pd_\gamma^2
+\Big(\dfrac{\cosh \eta}{\sinh \eta}-\dfrac{\sinh \eta}D \Big)
\pd_\eta -\dfrac{\sin \gamma}D\pd_\gamma \Big]\beta  \nn\\
&+2\Big(\pd_\eta f\pd_\eta \beta +\pd_\gamma
f\pd_\gamma \beta \Big)\dfrac{\cos f}{\sin f}  \nn\\
&-\dfrac{\alpha}{\kappa^2} \dfrac{D^2}{a^2} C=0,
\label{skeq4}
\eea
where
\bea
&A=\Big[n^2 K^2+\dfrac{m^2}{\sinh ^2\eta}
\Big((\pd_\eta f)^2+(\pd_\gamma f)^2\Big)\Big], \nn\\
&B = \Big\{n^2 \pd_\eta K\pd_\gamma \beta  
-n^2 \pd_\gamma K\pd_\eta \beta \nn\\
&+n^2 K\Big[\Big(\dfrac{\cosh \eta}{\sinh \eta} 
+\dfrac{\sinh \eta }D \Big)\pd_\gamma \beta  
-\dfrac{\sin \gamma}D \pd_\eta \beta \Big] \nn\\
&+\dfrac{m^2}{\sinh^2\eta}\Big[\partial_\eta^2
+\partial_\gamma^2 -\Big(\dfrac{\cosh \eta}{\sinh \eta}
-\dfrac{\sinh \eta}D\Big) \partial_\eta +\dfrac{\sin \gamma}D
\partial_\gamma \Big]f \Big\}, \nn\\
&C=\Big\{\pd_\eta K\pd_\gamma f -\pd_\eta f \pd_\gamma K  \nn\\
&+K\Big[\Big(\dfrac{\cosh \eta}{\sinh \eta}+\dfrac{\sinh \eta}D \Big)
\pd_\gamma -\dfrac{\sin \gamma}D\pd_\eta \Big)f\Big\}. \nn
\eea

With the ansatz (\ref{skkans}) the knot energy is given by
\bea
&E= {\sqrt \alpha} \kappa \Int \Big\{\dfrac{\kappa}{\sqrt \alpha}
\dfrac{a}{2D}\Big[(\pd_\eta f)^2+(\pd_\gamma f)^2   \nn\\
&+ \Big(n^2 \big((\pd_\eta \beta)^2 +(\pd_\gamma \beta)^2 \big)
+\dfrac{m^2}{\sinh^2 \eta}\Big) \sin^2 f \Big] \nn\\
&+\dfrac{\sqrt \alpha}{\kappa}\dfrac{D}{4a}\Big[n^2 K^2
+\dfrac{m^2}{\sinh ^2\eta} \Big((\pd_\eta f)^2   \nn\\
&+(\pd_\gamma f)^2\Big)\Big] \sin^2 f \Big\}
\sinh \eta d\eta d\gamma d\varphi.
\label{ske}
\eea
Minimizing the energy we reproduce the knot equation
(\ref{skeq4}), which confirms that ansatz (\ref{skkans}) is
consistent.

\begin{figure}
\includegraphics[scale=0.45]{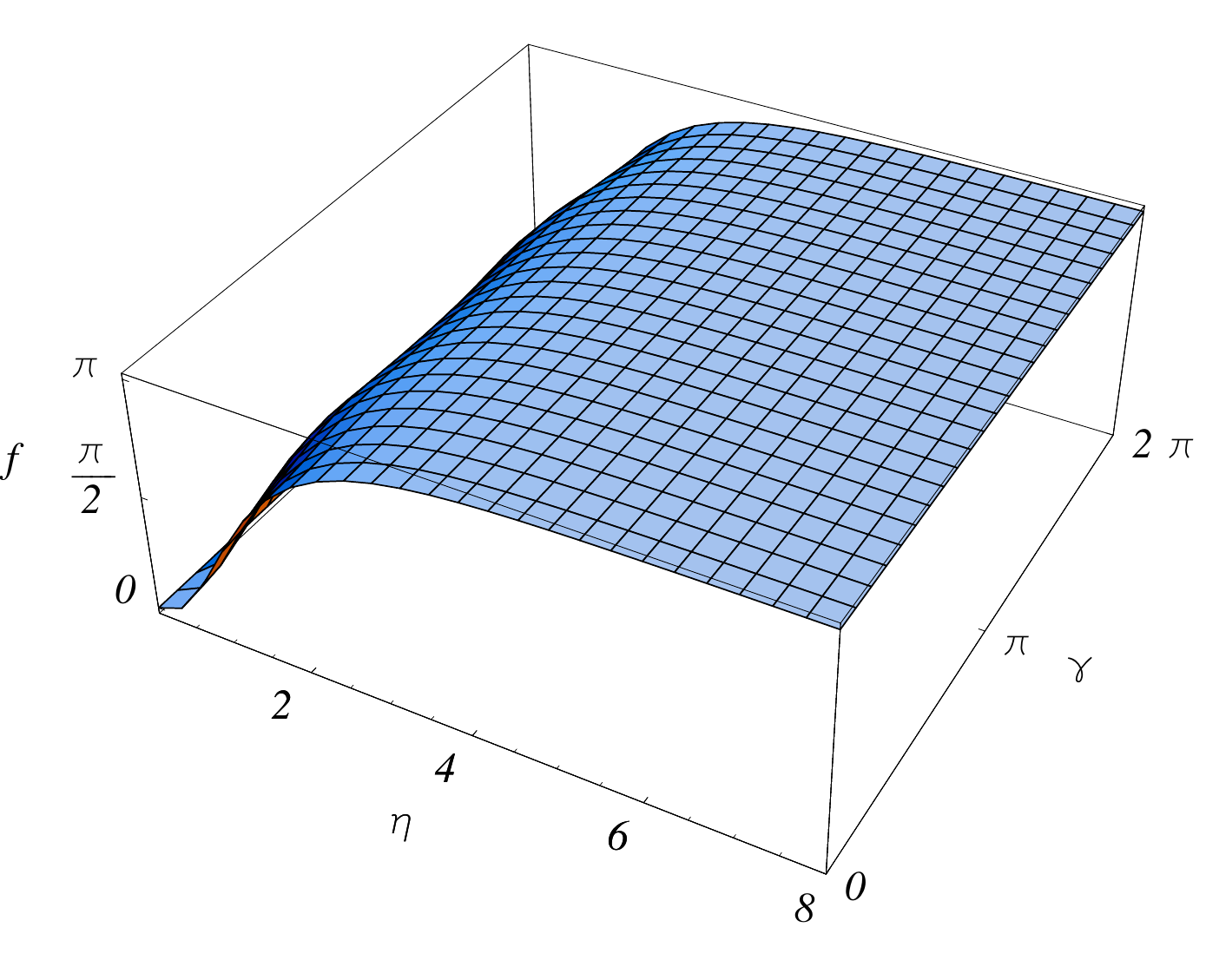}
\caption{(Color online). The $f$ configuration of the knot with
$m=n=1$ in Skyrme theory. Notice that here $\eta$ is dimensionless.}
\label{skkf}
\end{figure}

\begin{figure}
\includegraphics[scale=0.45]{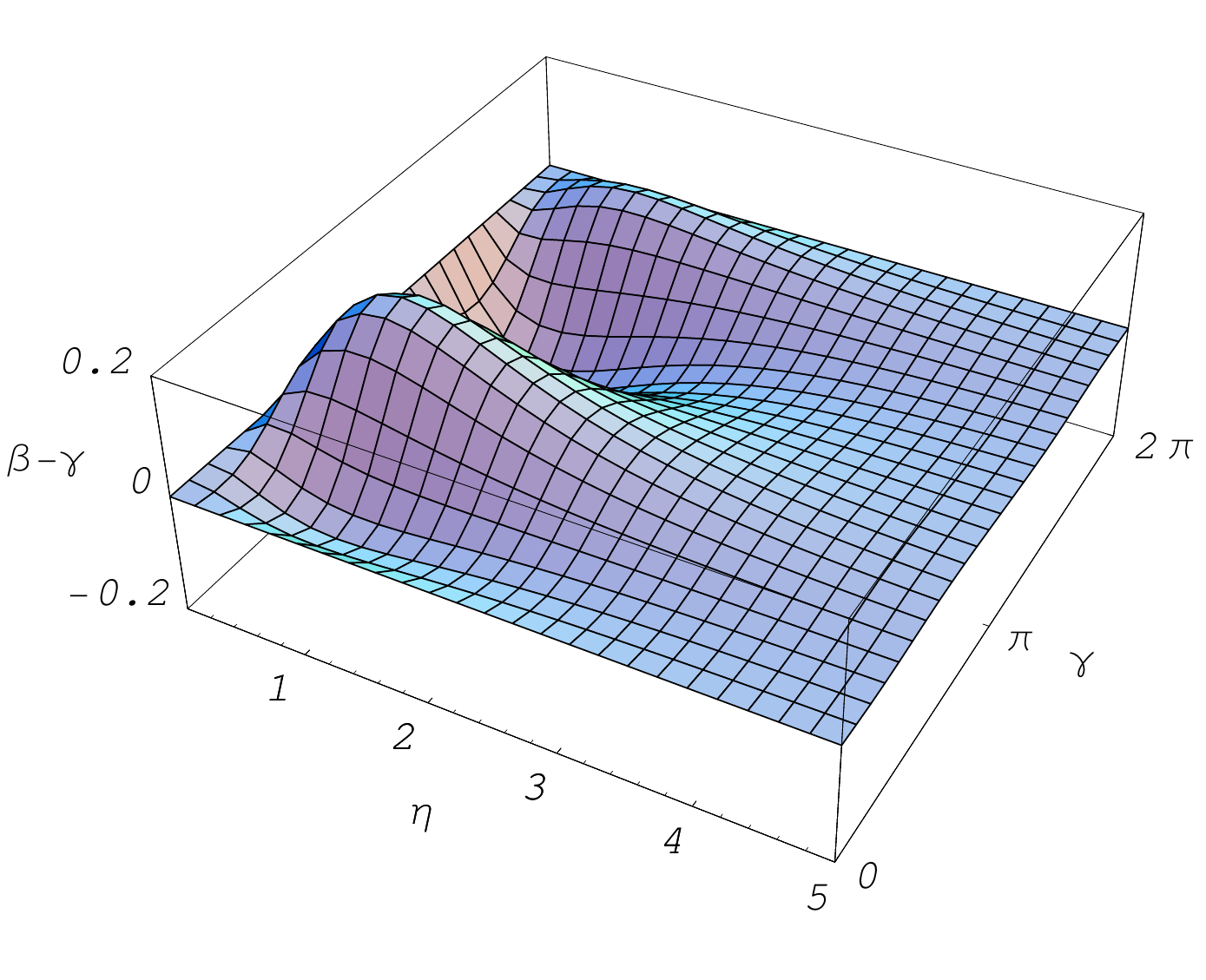}
\caption{(Color online). The $\beta$ configuration of the knot
with $m=n=1$ in Skyrme theory. Notice that here the actual
configuration shown is $\beta-\gamma$.}
\label{skkdo}
\end{figure}

In toroidal coordinates, $\eta=\gamma=0$  represents spatial
infinity and $\eta=\infty$ describes the torus center. So we
can try to obtain the knot solution imposing the following 
boundary condition
\begin{eqnarray}
&f(0,\gamma)=0,
~~~~~f(\infty,\gamma)=\pi, \nn\\
&\beta(\eta,0)=0,
~~~~~\beta(\eta,2 \pi)=2 \pi.
\label{knotbc}
\end{eqnarray}
Of course, an exact solution of (\ref{skeq4}) is extremely 
difficult to obtain, even numerically. In fact many known 
``knot solutions" are actually the energy profile of knots 
which minimizes the Hamiltonian \cite{fadd}. But here with 
the ansatz (\ref{skkans}) we can find an actual profile 
of the knot. For $m=n=1$ the knot radius $a$ which minimizes 
the energy is given by
\bea
a \simeq 1.21~\dfrac{\sqrt \alpha}{\kappa}.
\label{skkrad}
\eea
From this we obtain Fig. \ref{skkf} and Fig. \ref{skkdo} 
which shows the knot profile for $f$ and $\om$ of 
the lightest axially symmetric knot solution. 

With the numerical solution we can check the topology of 
the knot. From the ansatz (\ref{skkans}) we have the knot 
number
\bea
&Q_k=\dfrac{1}{32\pi^2}\Int \epsilon_{ijk}~C_i~H_{jk} d^3x \nn\\
&=\dfrac{mn}{8\pi ^2}\Int K \sin f d\eta d\gamma d\varphi
= \dfrac{mn}{4\pi} \Int \sin f df d\beta \nn\\
&= mn,
\label{kqn}
\eea
where the last equality comes from the boundary condition 
(\ref{knotbc}). This assures that our ansatz describes
the correct knot topology. Notice this is formally 
identical to the Chern-Simon integral of the helicity 
of the electromagnetic knot shown in (\ref{hel}).  

To clarify the meaning of (\ref{kqn}) we now calculate 
the magnetic flux of the knot. The helical magnetic field 
has two magnetic fluxes, $\Phi_{\hat \gamma}$ passing 
through the knot disk of radius $a$ in the $xy$-plane 
and $\Phi_{\hat \varphi}$ moving along the knot ring of 
radius $a$, given by
\begin{eqnarray}
&\Phi_{\hat{\gamma}} = \Int_{\gamma=\pi} H_{\hat{\gamma}}
\dfrac{a^2\sinh \eta}{D^2}d\eta d\varphi  \nn\\
&=m \Int_{\gamma=\pi} \sin f\pd_\eta f d\eta d\varphi
=4\pi m, \nn\\
&\Phi_{\hat{\varphi}} = \Int H_{\hat{\varphi}}
\dfrac{a^2}{D^2}d\eta d\gamma
=n \Int K \sin fd\eta d\gamma \nn\\
&=4\pi n.
\label{kflux}
\end{eqnarray}
This confirms the followings. First, the flux is quantized 
in the unit of $4\pi$. Second, the two fluxes are linked,
whose linking number is given by $mn$. This is precisely 
the knot number (\ref{kqn}). This provides the physical 
interpretation of the topological knot. 

The supercurrent which generates the twisted magnetic flux of
the knot is given by the conserved current density
\bea
&j_\mu =\dfrac{nD^2}{a^2}\Big(\pd_\eta +\dfrac{\cosh \eta}
{\sinh \eta}+\dfrac{\sinh \eta}D\Big) K \pd_\mu \gamma \nn\\
&+\dfrac{mD^2}{a^2}\Big[\Big(\pd_\eta-\dfrac{\cosh \eta }{\sinh \eta}
+\dfrac{\sinh \eta }D\Big)\sin f\pd_\eta f \nn\\
&+\Big(\pd_\gamma +\dfrac{\sin \gamma}D\Big)\sin f
\pd_\gamma f\Big] \pd_\mu \varphi.
\label{kcd}
\eea
From this we have two quantized supercurrents, $i_{\hat\gamma}$ 
which flows through the knot disk and $i_{\hat\varphi}$ which 
flows along the knot,
\bea
&i_{\hat{\gamma}} =\Int_{\gamma =\pi} j_{\hat{\gamma}}(\eta,\gamma)
\dfrac{a^2\sinh \eta}{D^2}d\eta d\varphi  \nn\\
&=\dfrac n{a}\Int_{\gamma =\pi} D\sinh \eta \Big(\pd_\eta
+\dfrac{\cosh \eta}{\sinh \eta}+\dfrac{\sinh \eta}D\Big)
K d\eta d\varphi,  \nn\\
&i_{\hat{\varphi}}
=\Int j_{\hat{\varphi}}\dfrac{a^2}{D^2}d\eta d\gamma  \nn\\
&=\dfrac m{a}\Int \dfrac D{\sinh \eta}
\Big[\Big(\pd_\eta -\dfrac{\cosh \eta}{\sinh \eta} 
+\dfrac{\sinh \eta}D\Big)\sin f\pd_\eta f \nn\\
&+\Big(\pd_\gamma +\dfrac{\sin \gamma}D \Big)
\sin f\pd_\gamma f \Big] d\eta d\gamma.
\eea
For $m=n=1$ we find numerically
\bea
&i_{\hat{\gamma}} \simeq \dfrac{23.9}{a},
~~~~~i_{\hat{\varphi}} \simeq 0.
\eea
Notice that $i_{\hat{\varphi}}$ is vanishing, but the current 
density $j_{\hat{\varphi}}$ is non-trivial. This tells that 
the knot has a net angular momentum around the symmetric axis 
which stablizes the knot.

\begin{figure}
\includegraphics[scale=0.5]{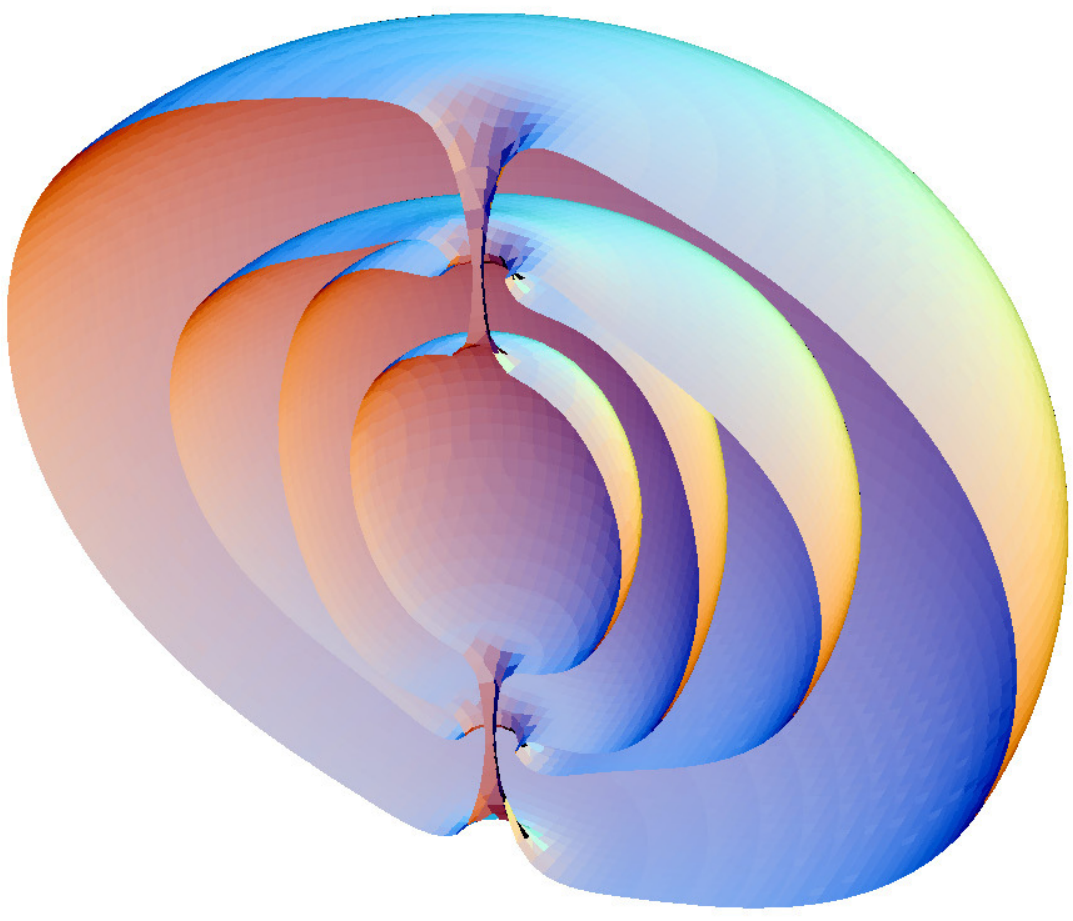}
\caption{(Color online). The 3-dimensional energy profile of 
the lightest axially symmetric knot with $m=n=1$ in Skyrme theory. 
Here the scale is in the unit of ${\sqrt \alpha}/\kappa$.}
\label{skk3d}
\end{figure}

One might wonder how big is the knot. The numerical result 
suggests that the radius of the vortex (the thickness of the knot) 
$r_0$ is roughly given by
\bea
r_0 \simeq a~{\rm csch}~2 \simeq
\dfrac{1}{3} \dfrac{\sqrt \alpha}{\kappa}.
\label{vrad}
\eea
This shows that the radius of the vortex ring is about 3.6 
times the radius of the vortex. From this we can construct 
a 3-dimensional energy profile of the knot. This is shown in 
Fig. \ref{skk3d}.

In mathematics the knot is described by the linking number 
of the preimage of the Hopf mapping. When the preimages 
of two points of the target space are linked, the mapping 
defines a knot. And the knot number of $\pi_3(S^2)$ 
is given by the linking number of two preimages fixed by 
the Chern-Simon index (\ref{kqn}) of the potential $C_\mu$.

The above result provides an alternative picture of knot. 
It shows that two real (physical) magnetic flux rings linked 
together makes the knot, whose knot number is given by 
the linking number of two flux rings. Clearly this is different 
from the above mathematical definition of knot based on 
the Hopf mapping. This is a dynamical manifestation of knot, 
the linking of two physical flux rings generated by 
dynamics \cite{plb04}.

Obviously two flux rings linked together can not be unlinked 
by any smooth deformation of the flux ring. This guarantees 
the topological stability of the knot. Moreover, this topological 
stability is backed up by the dynamical stability which comes 
from the dynamics of the flux ring. This is because the supercurrent 
which generates the quantized magnetic flux of the knot has 
two components, the one moving along the knot, and the other 
moving around the knot tube. And the supercurrent moving 
along the knot generates non-vanishing angular momentum 
around the $z$-axis, which provides the centrifugal force 
preventing the vortex ring to collapse. This is how the knot 
acquires the dynamical stability \cite{plb04}.

As we have argued, the knot is stable within the framework 
of Skyrme-Faddeev theory. However, we have to be cautious 
about the knot stability in Skyrme theory. This is because 
the Skyrme theory has the extra massless scalar field $\omega$ 
which (in principle) could destabilize the knot. So in Skyrme 
theory the proof of the stability of the knot becomes a non-trivial 
matter.

\section{Knots in Condensed Matter Physics}

The knots can also appear in two-component superfluid 
and two-gap superconductor \cite{pra05,prb06,epjb08}.
To see this let $\phi$ be a complex doublet which describes
a two-component Bose-Einstein condensate (BEC)
\bea
\phi = \dfrac {1}{\sqrt 2} \rho~\xi , ~~~~~(\xi^\dag \xi = 1)
\eea
and consider the following ``gauged" Gross-Pitaevskii type 
Lagrangian \cite{pra05}
\bea
&{\cal L} = - |D_\mu \phi|^2 - \dfrac {\lambda}{2}
\big(\phi^\dag \phi -\dfrac{\kappa^2}{\lambda} \big)^2
- \dfrac {1}{4} F_\mn^2,
\label{beclag}
\eea
where $D_\mu =  \pro_\mu + i g A_\mu$, $\kappa^2$ and $\lambda$ 
are the coupling constants. Of course, since we are interested 
in a neutral condensate, we identify the potential $A_\mu$ with 
the velocity field of $\xi$ 
\bea
g A_\mu = -i \xi^\dag \pro_\mu \xi.
\label{vpot}
\eea
With this the Lagrangian (\ref{beclag}) is reduced to
\bea
&{\cal L} = -\dfrac {1}{2} (\pro_\mu \rho)^2
- \dfrac {\rho^2}{2} \Big(|\pro_\mu
\xi |^2 - |\xi^\dag \pro_\mu \xi|^2 \Big) \nn\\
&-\dfrac{\lambda}{8} \Big(\rho^2 - \rho_0^2 \Big)^2
+ \dfrac {1}{4 g^2} (\pro_\mu \xi^\dag \pro_\nu \xi
- \pro_\nu \xi^\dag \pro_\mu \xi)^2,  \nn\\
&\rho_0^2=\dfrac{2\kappa^2}{\lambda}. 
\label{beclag1}
\eea
Notice that here the gauge field strength $F_\mn$ is replaced by
the non-vanishing vorticity of the velocity field (\ref{vpot}),
but the Lagrangian still retains the $U(1)$ gauge symmetry
of (\ref{beclag}).

From the Lagrangian we have the following equation of motion
\bea
& \pro^2 \rho - \Big(|\pro_\mu \xi |^2 - |\xi^\dag \pro_\mu \xi|^2 \Big)
\rho = \dfrac{\lambda}{2} (\rho^2 - \rho_0^2) \rho,\nn \\
&\Big\{(\pro^2 - \xi^\dag \pro^2 \xi) + 2 \Big(\dfrac {\pro_\mu
\rho}{\rho} - \xi^\dag \pro_\mu\xi \nn\\
&+ \dfrac {1}{g^2 \rho^2} \pro_\alpha (\pro_\mu
\xi^\dag \pro_\alpha \xi - \pro_\alpha \xi^\dag \pro_\mu \xi) \Big)
(\pro_\mu - \xi^\dag \pro_\mu \xi) \Big\} \xi \nn\\
&= 0.
\label{beceq1}
\eea
But remarkably, with
\bea
\hn = \xi^\dag \vec \sigma \xi,
\label{ndef}
\eea
we have
\bea
& (\pro_\mu \hn)^2 = 4 (|\pro_\mu \xi|^2
- |\xi^\dag \pro_\mu \xi|^2) \nn\\
&H_{\mu\nu} = \hn \cdot (\pro_\mu \hn \times \pro_\nu \hn)
= -2i (\pro_\mu \xi^\dag \pro_\nu \xi
- \pro_\nu \xi^\dag \pro_\mu \xi ) \nn\\
&= \pro_\mu C_\nu - \pro_\nu C_\mu.
\label{fmn}
\eea
This tells that the velocity potential (\ref{vpot}) plays
the role of the magnetic potential $C_\mu$ in (\ref{sfeq})
of Skyrme theory.

With (\ref{fmn}) the Lagrangian (\ref{beclag1})
is reduced to the following $CP^1$ Lagrangian
\bea
&{\cal L} = -\dfrac{1}{2}(\pro_\mu \rho)^2
-\dfrac {\rho^2}{2} (\pro_\mu \hn)^2
-\dfrac{\lambda}{8} (\rho^2-\rho_0^2)^2 \nn\\
&-\dfrac {1}{16g^2} (\pro_\mu \hn \times \pro_\nu \hn)^2.
\label{beclag3}
\eea
Furthermore the equation (\ref{beceq1}) can
be put into the form
\bea &\pro^2 \rho - \dfrac{1}{4} (\pro_i
\hn)^2 \rho = \dfrac{\lambda}{2} (\rho^2 - \rho_0^2)
\rho, \nn \\
&\hn \times \pro^2 \hn + 2 \dfrac{\pro_i \rho}{\rho} \hn 
\times \pro_i \hn + \dfrac{1}{g^2 \rho^2} \pro_i H_{ij} \pro_j \hn = 0.
\label{beceq2}
\eea
The reason why we can express (\ref{beceq1}) completely
in terms of $\hn$ (and $\rho$) is that the Abelian gauge 
invariance of (\ref{beclag}) effectively reduces the target 
space of $\xi$ to the gauge orbit space $S^2 = S^3/S^1$, 
which is identical to the target space of $\hn$.

This analysis clearly shows that the above theory of
two-component BEC is closely related to the Skyrme theory.
In fact, in the vacuum $\rho^2=\rho_0^2$ the Lagrangian 
reduces to the Skyrme-Faddeev Lagrangian
\bea
&{\cal L} = -\dfrac {\rho_0^2}{2} \Big(|\pro_\mu
\xi |^2 - |\xi^\dag \pro_\mu \xi|^2 \Big)  \nn\\
&-\dfrac {1}{4g^2} (\pro_\mu \xi^\dag \pro_\nu \xi
-\pro_\nu \xi^\dag \pro_\mu \xi)^2 \nn\\
&=-\dfrac {\rho_0^2}{2} (\pro_\mu \hn)^2
-\dfrac {1}{16g^2} (\pro_\mu \hn \times \pro_\nu \hn)^2.
\label{beclag2}
\eea
This shows that the three Lagrangians (\ref{sflag}), (\ref{mrcd1}),
and (\ref{beclag2}) are all identical to each other, which 
confirms that the Skyrme theory and the theory of 
two-component superfluid are indeed very similar. This 
implies that the Skyrme-Faddeev theory could also
be regarded as a theory of two-component superfluid.

\begin{figure}[t]
\includegraphics[height=4cm, width=7cm]{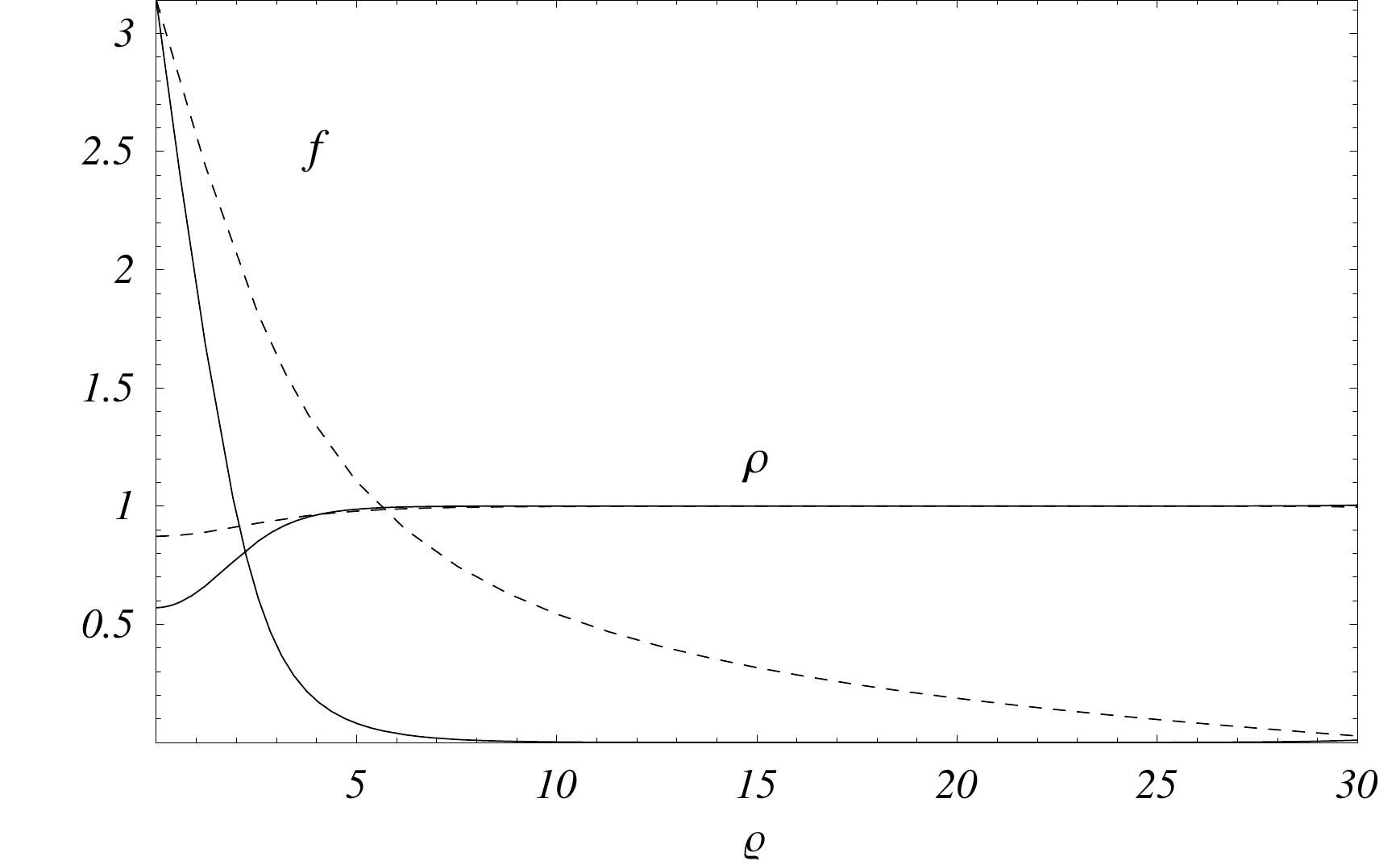}
\caption{The non-Abelian vortex (dashed line) with $m=0,n=1$ and
the helical vortex (solid line) with $m=n=1$ in the gauge theory
of two-component BEC. Here we have put $\lambda/g^2=1$,
$k=0.64~{\sqrt \lambda}\rho_0$, and $\varrho$ is in the unit of
$1/{\sqrt \lambda}\rho_0$.}
\label{becheli}
\end{figure}

The above observation also reveals two important facts.
First, it tells that the Skyrme theory has a $U(1)$ gauge 
symmetry as we have demonstrated in (\ref{skyu1}). 
Indeed, it is this gauge symmetry which allows us to 
establish the existence of the Meissner effect in Skyrme 
theory. Second, it provides a new meaning to $H_\mn$. 
The two-form now describes the vorticity of the velocity 
field of the superfluid $\xi$. In other words, the non-linear 
Skyrme interaction can be interpreted as the vorticity 
interaction of the superfluid. It is well-known that 
the vorticity plays an important role in superfluid \cite{ho}. 
But creating a vorticity in superfluid costs energy. So it 
makes a perfect sense to include the vorticity interaction 
in the theory of superfluid. And the Skyrme-Faddeev 
Lagrangian naturally contains this interaction.

Moreover, the above observation makes it clear that
the above gauge theory of two-component BEC 
has a knot very similar to the Faddeev-Niemi 
knot \cite{pra05,prb06}. Indeed with the following knot 
ansatz for two-component BEC
\bea
&\phi=\dfrac{1}{\sqrt 2} \rho (\eta, \gamma) \xi, \nn\\
&\xi=\left(\begin{array}{cc} 
\cos \dfrac{f(\eta,\gamma)}{2} \exp (-im\varphi )   \\
\sin \dfrac{f(\eta ,\gamma)}{2} \exp (in \beta(\eta,\gamma)) 
\end{array} \right),   
\label{bkans}
\eea
we have
\bea
&\hn= \xi^{\dagger} \vec \sigma \xi
=\left(\begin{array}{ccc} 
sin f \cos (n\beta+m\varphi) \\
\sin f \sin (n\beta + m\varphi) \\
\cos f 
\end{array} \right),   \nn\\
&C_\mu = -2i \xi^{\dagger} \partial _\mu \xi \nn\\
&= n(\cos f -1) \partial _\mu \beta
+m(\cos f +1)\partial _\mu \varphi. \nn
\label{becn}
\eea
Clearly this is identical to the Faddeev-Niemi knot 
ansatz shown in (\ref{skkans}). With the ansatz (\ref{bkans}) 
we can obtain a knot solution very similar to 
the Faddeev-Niemi knot \cite{pra05,prb06}. This is 
shown in Fig. \ref{becheli}. The only difference between 
the two knots is that the one in BEC has a dressing 
of an extra scalar field $\rho$ which represents 
the degree of the condensation. 

Clearly the knot in two component BEC describes a vorticity 
knot. And here the complex doublet $\xi$ provides the knot 
topology, 
\bea
&Q_k = \dfrac {1}{4\pi^2} \Int \epsilon_{ijk} \xi^{\dagger}
\pd_i \xi ( \pd_j \xi^{\dagger} \pd_k \xi ) d^3 x  \nn\\
&= \dfrac{1}{32\pi^2} \Int \epsilon_{ijk} C_i H_{jk} d^3x.
\label{bkqn}
\eea
This is the Chern-Simon index which is mathematically 
identical to the helicity (\ref{hel}) of the electromagnetic 
knot.
  
But $\xi$ defines the mapping $\pi_3(S^3)$ from 
the compactified space $S^3$ to the target space $S^3$.
Apparently this is the baryon topology of the skyrmion, 
not the knot topology $\pi_3(S^2)$. This has led to 
a misleading statement in the literature that the vorticity 
knot can be identified as a skyrmion \cite{batt}. But 
this is wrong. The reason is that the target space 
has the Hopf fibering $S^3\simeq S^2\times S^1$, and 
the hidden $U(1)$ gauge symmetry in BEC effectively 
reduces the target space to $S^2$, so that $\pi_3(S^3)$ 
defined by $\xi$ actually reduces to $\pi_3(S^2)$. So 
actually the knot topology here is described by $\hn$
given by (\ref{becn}). In comparoson in Skyrme theory
the scalar field $\om$ describes the $S^1$ fiber, 
so that the baryon topology $\pi_3(S^3)$ can not be 
reduced to $\pi_3(S^2)$. 

We can have similar knot in multi-gap 
superconductors \cite{prb06,epjb08}. In the above BEC 
the gauge interaction was a self-induced interaction. 
But when the doublet is charged, the gauge interaction 
can be treated as independent. In this case the theory 
can describe a two-gap superconductor. But even in 
this case the knot topology and thus the knot itself 
should survive. This implies that two-gap superconductor 
could also have a topological knot.

The knot in two-gap superconductor could be either relativistic
or non-relativistic, and appear in both Abelian and non-Abelian
setting \cite{pra05,prb06}. In this paper we will discuss 
the relativistic knot in the Abelian setting (a non-relativistic 
Gross-Pitaevskii type theory gives an identical result). Consider 
a charged doublet scalar field $\phi$ coupled to real electromagnetic 
field,
\bea
&{\cal L} = - |D_\mu \phi|^2 + \kappa^2 \phi^{\dagger}\phi 
- \dfrac{\lambda}{2} (\phi^{\dagger} \phi)^2
- \dfrac{1}{4} F_\mn^2 , \nn\\
&D_\mu \phi = (\partial_\mu - ig A_\mu) \phi.
\label{sclag}
\eea
The Lagrangian has the equation of motion
\bea
&D^2\phi =\lambda(\phi^{\dagger} \phi
-\dfrac{\kappa^2}{\lambda})\phi, \nn\\
&\partial_\mu F_\mn = j_\nu =  i g \Big[(D_\nu
\phi)^{\dagger}\phi - \phi ^{\dagger}(D_\nu \phi) \Big].
\label{sceq1}
\eea
Now, with
\bea
\phi =\dfrac{1}{\sqrt 2} \rho \xi,~~~~~{\xi}^{\dagger}\xi = 1,
~~~~~\hat n = \xi^{\dagger} \vec \sigma \xi,
\eea
we can reduce (\ref{sceq1}) to 
\bea
&\partial ^2 \rho - \Big(\dfrac{1}{4} (\pd_\mu \hn)^2
+ g^2 (A_\mu + \tilde A_\mu)^2 \Big) \rho   \nn\\
&= \dfrac{\lambda}{2} (\rho^2-\rho_0^2)\rho, \nn\\
&\hn \times \pd^2 \hn + 2 \dfrac{\pd_\mu \rho}{\rho} 
\hn \times \pd_\mu \hat n
+ \dfrac{2}{g\rho^2} \pd_\mu F_\mn \pd_\nu \hn =0, \nn\\
&\pd_\mu F_\mn = j_\mu =g^2 \rho^2 (A_\mu + \tilde A_\mu), \nn\\
&\tilde A_\mu=-\dfrac{i}{g} \xi^{\dagger}\pd_\mu \xi,
~~~~~\rho_0=\dfrac{2\kappa^2}{\lambda}.
\label{sceq2}
\eea
This is the equation for two-gap superconductor.
Notice that with $A_\mu=-\tilde A_\mu$ the first two 
equations reduce to (\ref{beceq2}). This tells that 
the gauge theory of two-component BEC and the above 
theory of two-gap superconductor are closely related.

To obtain the desired knot we first construct a superconducting
helical magnetic vortex. Let 
\bea
&\rho=\rho(\varrho), 
~~~~~\xi = \left(\begin{array}{cc} 
\cos \dfrac{f(\varrho)}{2} \exp (-in\varphi) \\
\sin \dfrac{f(\varrho)}{2} \exp (imkz) \end{array} \right), \nn\\
&A_\mu= \dfrac{1}{g} \big(n A_1(\varrho) \partial_\mu\varphi
+ mk A_2(\varrho) \partial_\mu z \big), \nn\\
&\n= \xi^\dag \vec \sigma \xi
=\left(\begin{array}{ccc}  \sin{f(\varrho)}\cos{(n\varphi+mkz)} \\
\sin{f(\varrho)}\sin{(n\varphi+mkz)} \\ 
\cos{f(\varrho)} \end{array} \right), \nn\\
&\tilde A_\mu=-\dfrac{n}{2g} \big(\cos{f(\varrho)}+1\big)
\partial_\mu \varphi \nn\\
&- \dfrac{mk}{2g} \big(\cos{f(\varrho)}-1\big) \partial_\mu z.
\label{scans}
\eea
With this we have
\bea
&j_\mu = g\rho^2 \Big(n \big(A_1-\dfrac{\cos{f}+1}{2}\big)
\partial_\mu \varphi \nn\\
&+ mk\big(A_2-\dfrac{\cos{f}-1}{2}\big)
\partial_\mu z \Big),
\label{sc}
\eea
and (\ref{sceq2}) becomes
\bea
&\ddot{\rho}+\dfrac{1}{\varrho}\dot\rho
- \Big[\dfrac{1}{4} \Big(\dot{f}^2
+\big(\dfrac{n^2}{\varrho^2} + m^2 k^2 \big) \sin^2{f}\Big) \nn\\
&+\dfrac{n^2}{\varrho^2} \Big(A_1-\dfrac{\cos{f}+1}{2}\Big)^2
+ m^2 k^2 \Big(A_2-\dfrac{\cos{f}-1}{2}\Big)^2 \Big]\rho \nn\\
&= \dfrac{\lambda}{2}(\rho^2-\rho_0^2)\rho, \nn\\
&\ddot{f} + \big(\dfrac{1}{\varrho}
+2\dfrac{\dot{\rho}}{\rho} \big)\dot{f}
-2 \Big(\dfrac{n^2}{\varrho^2}\big(A_1-\dfrac{1}{2}\big) \nn\\
&+ m^2 k^2 \big(A_2+\dfrac{1}{2}\big) \Big)\sin{f} = 0, \nn\\
&\ddot{A}_1-\dfrac{1}{\varrho}\dot{A}_1 -g^2 \rho^2
\Big(A_1-\dfrac{\cos{f}+1}{2}\Big) = 0, \nn\\
&\ddot{A}_2+\dfrac{1}{\varrho}\dot{A}_2 -g^2 \rho^2
\Big(A_2-\dfrac{\cos{f}-1}{2}\Big) = 0.
\label{sceq3}
\eea
Now, we impose the following boundary condition for 
the non-Abelian vortices,
\bea
&\rho (0) = 0,~~~\rho(\infty) = \rho_0, 
~~~f (0) = \pi,~~~f (\infty) = 0, \nn\\
& A_1 (0) = -1,~~~A_1 (\infty) =1.
\label{scbc}
\eea
This need some explanation, because the boundary
value $A_1(0)$ is chosen to be $-1$, not $0$. This is to
assure the smoothness of $\rho(\varrho)$ and $f(\varrho)$
at the origin. Only with this boundary value they
become analytic at the origin.

One might object the boundary condition, because it creates 
an apparent singularity in the gauge potential at the origin. 
But this singularity is an unphysical (coordinate) singularity 
which can easily be removed by the gauge transformation
\bea
\phi \rightarrow \phi \exp(in\varphi),
~~~~A_\mu \rightarrow A_\mu + \dfrac{n}{g} \pd_\mu \varphi,
\eea
which changes the boundary condition $A_1(0)=-1$ and 
$A_1(\infty)=1$ to $A_1(0)=0$ and $A_1(\infty) =2$.
Mathematically this boundary condition has a deep origin, 
which has to do with the fact that the Abelian $U(1)$ 
runs from $0$ to $2\pi$, but the $S^1$ fiber of $SU(2)$ 
runs from $0$ to $4\pi$. As for $A_2(\varrho)$, we 
choose $A_2(\infty)=0$ to make the supercurrent vanishing 
at infinity and require the vortex superconducting. As 
we will see, this requires a logarithmic divergence for 
$A_2(0)$. The boundary condition will have an important 
consequence in the following.

\begin{figure}
\includegraphics[width=7cm, height=4cm]{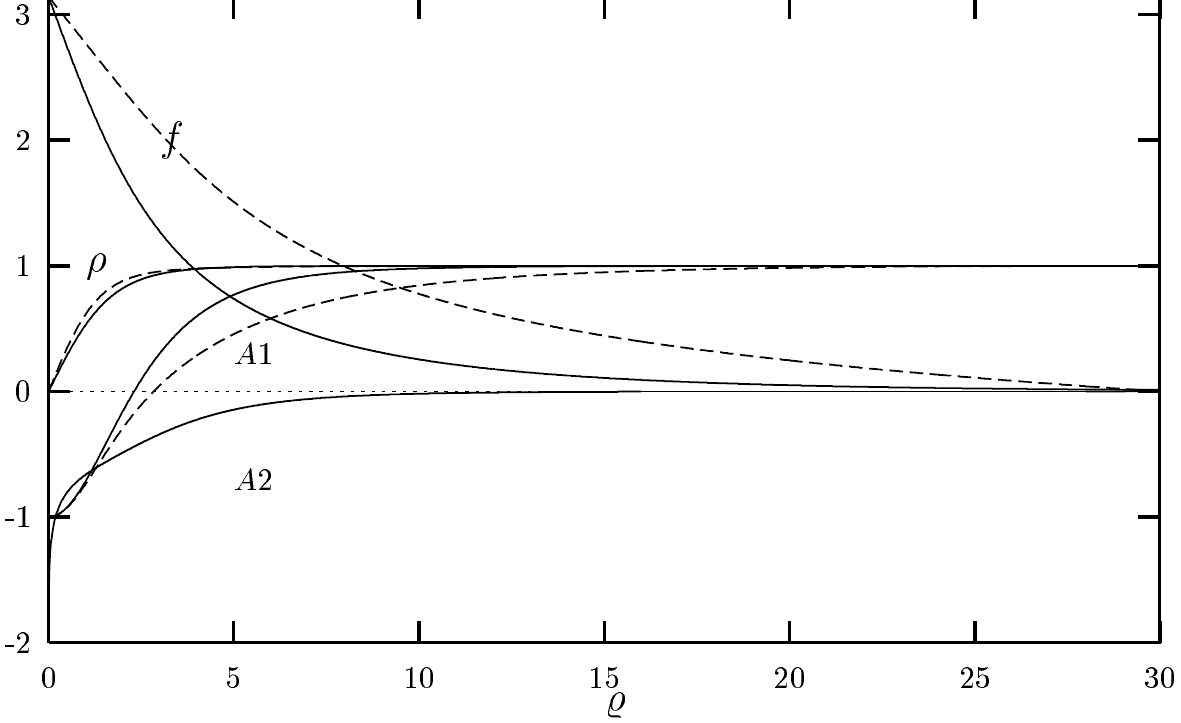}
\caption{The non-Abelian vortex (dashed line) with
$m=0,n=1$ and the helical vortex (solid line) with $m=n=1$ in
two-gap superconductor. Here we have put $g=1,~\lambda=2$,
$k=\rho_0/10$, and $\varrho$ is in the unit of $1/\rho_0$.
Notice that $A_2$ has a logarithmic singularity at the origin.}
\label{helisc}
\end{figure}

With the boundary condition we can integrate (\ref{sceq3})
and obtain the non-Abelian vortex solution of the two-gap 
superconductor. This is shown in Fig. \ref{helisc}. 
The solution is very similar to the one we have in 
two-component BEC. When $m=0$, the solution (with 
$A_2=0$) describes an untwisted non-Abelian vortex. 
But when $m$ is not zero, it describes a helical magnetic 
vortex periodic in $z$-coordinate. Notice that at the core
the vortex starts from the second component, but 
at the infinity the first component takes over completely.
This is due to the boundary condition $f(0)=\pi$ and 
$f(\infty)=0$, which assures that our solution describes 
a genuine non-Abelian vortex. This is true even when 
$m=0$. Only when $f=0$ (or $f=\pi$) the doublet 
effectively becomes a singlet, and (\ref{sceq3}) describes 
the Abelian Abrikosov vortex of one-gap superconductor.

There are important differences between the non-Abelian 
vortex and the Abrikosov vortex. First the non-Abelian 
vortex has a non-Abelian magnetic flux quantization. 
Indeed the quantized magnetic flux $\hat \phi_z$ of 
the non-Abelian vortex along the $z$-axis is given by 
\bea
&H_z= \dfrac{n}{g} \dfrac{\dot A_1}{\varrho}, \nn\\
&\hat \phi_z = \Int H_z d^2x
= \dfrac{2\pi n}{g} \big[A_1(\infty) - A_1(0)\big] \nn\\
&= \dfrac{4\pi n}{g}.
\label{zflux}
\eea
Notice that the unit of the non-Abelian flux is $4\pi/g$,
not $2\pi/g$. This is a direct consequence of the boundary 
condition (\ref{scbc}). 

This non-Abelian quantization of magnetic flux comes
from the non-Abelian topology $\pi_2(S^2)$ of the doublet 
$\xi$, or equivalently the triplet $\hn$, whose topological
quantum number is given by
\bea
&q = \dfrac {g}{4\pi } \Int H_z d^2 x
= - \dfrac {1}{4\pi} \Int \epsilon_{ij} \pd_i \xi^{\dagger}
\pd_j \xi d^2 x \nn\\
&= \dfrac {1}{8\pi} \Int \epsilon_{ij} \hn \cdot (\pd_i \hn
\times \pd_j \hn) d^2 x = n.
\label{scqn}
\eea
This distinguishes our non-Abelian vortex from
the Abelian vortex whose topology is fixed by $\pi_1(S^1)$.

Another important feature of the non-Abelian vortex 
is that it carries a non-vanishing supercurrent
along the $z$-axis,
\bea
&i_z = mkg \Int \rho^2 \big(A_2
-\dfrac{\cos f +1}{2}\big)\varrho d\varrho d\varphi \nn\\
&= \dfrac{2\pi mk}{g} \Int \big(\ddot A_2
+\dfrac{1}{\varrho} \dot A_2 \big) \varrho d\varrho \nn\\
&= \dfrac{2\pi mk}{g} (\varrho \dot A_2) \Big|_{\varrho=0}^{\varrho=\infty}
=-\dfrac{2\pi mk}{g} (\varrho \dot A_2) \Big|_{\varrho=0}.
\label{scz}
\eea
This is due to the logarithmic divergence of $A_2$ at the origin.

Notice that the superconducting helical vortex has only 
a heuristic value, because one needs an infinite energy 
to create it (since the magnetic flux around the vortex 
becomes divergent because of the singularity of $A_2$ 
at the origin). With the helical vortex, however, one can 
make a vortex ring by smoothly bending and connecting
two periodic ends. In the vortex ring the infinite magnetic
flux of $A_2$ can be made finite making the finite 
supercurrent (\ref{scz}) of the vortex ring produce a finite 
flux, and we can fix the flux to have the value $4\pi m/g$
by adjusting the current with $k$. With this the vortex ring 
now becomes a topologically stable knot.

To see this notice that the doublet $\xi$, after forming a knot,
acquires a non-trivial topology $\pi_3(S^2)$ which provides
the knot number,
\bea
&Q = - \dfrac {1}{4\pi^2} \Int \epsilon_{ijk} \xi^{\dagger}
\pd_i \xi ( \pd_j \xi^{\dagger} \pd_k \xi ) d^3 x \nn\\
&= \dfrac{g^2}{32\pi^2} \Int \epsilon_{ijk} C_i
(\pd_j C_k - \pd_k C_j) d^3x=mn.
\label{sckqn}
\eea
Again this is nothing but the Chern-Simon index of the potential
$C_\mu$, which is mathematically identical to the helicity 
of the electromagnetic knot (\ref{hel}) we discussed before. 
This tells that our knot is also made of two quantized magnetic 
flux rings linked together whose knot number is fixed by
the linking number $mn$. Obviously two flux rings linked 
together can not be separated by any continuous deformation of
the field configuration. This provides the topological stability
of the knot.

Again this topological stability is backed up by a dynamical 
stability. To see this notice that the supercurrent
of the knot has two components, the one around the knot tube 
which confines the magnetic flux along the knot, but more 
importantly the other along the knot which creates a magnetic 
flux passing through the knot disk. This component of 
supercurrent along the knot now generates a net angular 
momentum which provides the centrifugal repulsive force 
preventing the knot to collapse. This makes the knot 
dynamically stable.

To compare our knot with the Abrikosov vortex ring (made of 
the Abrikosov vortex in conventional superconductor),
notice that the Abrikosov knot is empty (i.e., does not 
carry a net supercurrent). As importantly it is unstable, 
and collapses immediately.

In contrast our knot has a helical supercurrent, and is 
stable. Furthermore these two features are deeply related. 
The helical supercurrent plays a crucial role to stablize 
the vortex ring by providing the net angular momentum, 
which prevents the collapse of the vortex ring. And this 
helical supercurrent originates from the knot topology. 
This remarkable interplay between topology and dynamics 
is what provides the stability of the knot. The nontrivial 
topology expresses itself in the form of the helical 
supercurrent, which in turn provides the dynamical 
stability of the knot. We emphasize that this supercurrent 
is what distinguishes our knot from the Abrikosov vortex 
ring, which has neither topological nor dynamical stability.

\section{Discussions}

Topology and topological objects have been playing increasingly 
important roles in physics, and certainly will play more important 
roles in the future. In this paper we have discussed the impact 
of topological objects, in particular the knots, in physics. As we 
have seen, the knots appear almost everywhere in physics, 
atomic physics, condensed matter physics, nuclear physics, 
high energy physics, even in gravitation.

The Skyrme theory provides an ideal platform for us to study 
the topological objects in physics. Originally it was proposed 
as a low energy effective theory of strong interaction where 
the baryons appear as the topological solitons made of 
pions, the Nambu-Goldstone field of the chiral symmetry 
breaking. But the Skyrme theory has many faces. As we have 
pointed out, the theory can actually be viewed as a theory of 
monopole which has the built-in Meissner effect. But from 
the topological point of view the most important feature of 
the theory is that it has almost all topological objects that 
we can find in physics.

It has the string (the baby skyrmion) which has the $\pi_1(S^1)$ 
topology, the Wu-Yang monopole which has the $\pi_2(S^2)$ 
topology, the skyrmion which has the $\pi_3(S^3)$ topology, and 
the Faddeev-Niemi knot which has the $\pi_3(S^2)$ topology. 
As importantly, the monopole plays the crucial role in all these 
objects. The string can be viewed as the magnetic vortex made 
of monopole-antimonopole pair infinitely separated apart, 
the skyrmion can be viewed as the regularized monopole 
which has the finite energy, and the knot can be viewed as 
the twisted magnetic vortex ring.   

Moreover, we have shown that the skyrmions actually carry 
the monopole number. So they are classified by two topological 
numbers, the baryon number and the monopole number. 
Furthermore, we have shown that the baryon number could 
be replaced by the radial (shell) number, so that the skyrmions 
can be classified by two integers $(m,n)$, the monopole 
number $m$ which describes the $\pi_2(S^2)$ topology 
of the $\hn$ field and the radial (shell) number $n$ 
which describes the $\pi_1(S^1)$ topology of the $\om$ 
field. In this scheme the baryon number $B$ is given 
by the product of two integers $B=mn$. This comes 
from the following observations. First, the SU(2) space $S^3$ 
admits the Hopf fibering $S^3\simeq S^2\times S^1$. 
Second, the Skyrme theory has two variables, the angular 
variable $\om$ which represents the $\pi_1(S^1)$ topology 
and the $CP^1$ variable $\hn$ which represents 
the $\pi_2(S^2)$ topology.

But the most important feature of the Skyrme theory for 
our discussion is that it has the prototype knot which 
can be interpreted as a twisted magnetic vortex ring
in which the knot number is given by the linking number 
of two quantized magnetic fluxes. What is remarkable 
is that this knot is made of real magnetic flux whose 
linking number describes the knot topology. This teaches 
us how to construct the knot, twisting the magnetic 
vortex to make it periodic along the vortex and making 
it a vortex ring connecting the two periodic ends 
together.

We can construct similar knots using this method in 
condensed matters. In two-component BEC we can 
obtain the vorticity knot twisting the vorticity vortex 
and making it a twisted vorticity ring. Similarly in two-gap 
superconductor we can have the magnetic vortex knot 
twisting the magnetic vortex and making it a twisted 
magnetic vortex ring.   

But perhaps the most interesting knot in physics is 
the electromagnetic knot in Maxwell's theory which 
can be viewed as electromagnetic geon. This knot 
has all features of an elementary particle, and behaves 
like an elementary particle with descrte mass and spin, 
in spite of the fact that it is a classical object. This is 
astonishing. This tells that the Wheeler's dream is 
not just a day dream. 

In this paper we have concentrated on knots, but before 
we close we like to emphasize the importance of topology 
in general in physics. Kelvin first understood the potential
importance of topology \cite{kelvin}. Of course, Kelvin's 
dream that topology could make the elementary particles 
(i.e., the atoms in his time) stable has not been realized yet. 
As we know, so far we have no elementary particle whose 
stability comes from topology. Nevertheless topology 
could play a fundamental role in physics as Dirac has 
taught us \cite{dirac}. In fact we could have a topologically 
stable elementary particle in the near future, the electroweak 
monopole.  

Ever since Dirac proposed his monopole, the monopole 
has been the most important topological object in physics.
There are two outstanding monopoles, Dirac monopole 
and 'tHooft-Polyakov monopole, but both have problems. 
The Dirac monopole is the singular Abelian monopole 
based on the $\pi_1(S^1)$ topology, which exists only 
when the U(1) bundle becomes non-trivial (i.e., when 
the U(1) bundle admits no global section). So it becomes 
optional (does not have to exist) in Maxwell's theory. 
More seriously, in the unification of electromagnetic 
and weak interactions it changes to the electroweak 
monopole \cite{plb97}. So, most probably the Dirac 
monopole may not exist in nature.  

This means that, if the standard model is correct, 
the electroweak monopole must exist. If so, the discovery 
of the electroweak monopole, not the Higgs particle, 
should be the final  and topological test of the standard 
model. And MoEDAL at CERN is actively searching 
for the monopole \cite{medal}. If discovered, it would 
be the first topologically stable elementary particle 
in human history. Without doubt, this would be the 
most dramatic evidence of the importance of topology 
in physics. The knot could have a similar impact in
physics in the future.   
 
{\bf ACKNOWLEDGEMENT}

~~~The work is supported in part by National Natural Science 
Foundation of China (Grant No. 11575254), National Research 
Foundation of Korea funded by the Ministry of Education 
(Grants 2015-R1D1A1A0-1057578), and by the Center for 
Quantum Spacetime at Sogang University.


\begin{thebibliography}{99}
\bibitem{kelvin} L. Kelvin, Trans. Roy. Soc. (Edington) {\bf 25}, 
217 (1868); P.G. Tait, Scientific Papers, {\bf 1}, 136 (1911).
\bibitem{dirac} P.A.M. Dirac, Proc. Roy. Soc. London, 
{\bf A133}, 60 (1931); Phys. Rev. {\bf 74}, 817 (1948).
\bibitem{hopf} H. Hopf, Math. Analen, {\bf 104}, 637 (1931).
\bibitem{bott} R. Bott and L.W. Tu, {\it Differential Forms in 
Algebraic Topology}, New York, (1982).

\bibitem{wu} T.T. Wu and C.N. Yang, Phys. Rev. {\bf D12}, 3845 (1975).
\bibitem{prl80} Y.M. Cho, Phys. Rev. Lett. {\bf 44}, 1115 (1980); 
Phys. Lett. {\bf B115}, 125 (1982).
\bibitem{thooft} G. 't Hooft, Nucl. Phys. {\bf B79}, 
276 (1974); A.M. Polyakov, JETP Lett. {\bf 20}, 194 (1974).
\bibitem{dokos} C. Dokos and T. Tomaras, Phys. Rev. {\bf D21}, 
2940 (1980).
\bibitem{cab} B. Cabrera, Phys.  Rev. Lett.  {\bf 48}, 1378~(1982). 
\bibitem{plb97} Y.M. Cho and D. Maison, Phys. Lett. {\bf B391}, 
360 (1997).
\bibitem{yang} Yisong Yang, Proc. Roy. Soc. London, {\bf A454}, 
155 (1998); Yisong Yang, {\it Solitons in Field Theory and 
Nonlinear Analysis} (Springer Monographs in Mathematics), 
p. 322 (Springer-Verlag) 2001.

\bibitem{bohm} Y. Aharonov and D. Bohm, Phys. Rev. {\bf 115}, 485 (1959).
\bibitem{berry} M.V. Berry, Proc. Roy. Soc. (London) {\bf A 392}, 45 (1984). 

\bibitem{sky} T.H.R. Skyrme, Proc. Roy. Soc. (London) {\bf 260}, 127
(1961); {\bf 262}, 237 (1961); Nucl. Phys. {\bf 31}, 556 (1962).
\bibitem{witt} G. Adkins, C. Nappi, and E. Witten,
Nucl. Phys. {\bf B228}, 552 (1983); A. Jackson and M. Rho, 
Phys. Rev. Lett. {\bf 51}, 751 (1983).
\bibitem{prep} See, for example, I. Zahed and G. Brown,
Phys. Rep. {\bf 142}, 1 (1986), and references therein.
\bibitem{man} N.S. Manton, Phys. Lett. {\bf B192}, 177 (1987).
\bibitem{bat} R.A. Battye and P.M. Sutcliffe, Phys. Rev. Lett. {\bf 79}, 
363 (1997); C.J. Houghton, N.S. Manton, and P.M. Sutcliffe,
Nucl. Phys. {\bf B510},507 (1998).

\bibitem{sut} R.A. Battye and P.M. Sutcliffe, Phys. Rev. Lett. 
{\bf 86}, 3989 (2001); Rev. Math. Phys. {\bf 14}, 29 (2002).
\bibitem{sky108} D.T.J. Feist, P.H.C. Lau, N.S. Manton, Phys. Rev. 
{\bf D87}, 085034  (2013).
\bibitem{hoyle} F. Hoyle, Astrophys. J. Suppl. {\bf 1}, 121 (1954).
\bibitem{epel} E. Epelbaum, H. Krebs, D. Lee, and U. Meissner,
Phys. Rev. Lett. {\bf 106}, 192501 (2011); 
P.H.C. Lau and N.S. Manton, Phys. Rev. Lett. {\bf 113}, 
232503 (2014); C.J. Halcrow and N.S. Manton, JHEP {\bf 1501}, 016 (2015).

\bibitem{fadd} L. Faddeev and A. Niemi, Nature {\bf 387}, 58 (1997);
R. Battye and P. Sutcliffe, Phys. Rev. Lett. {\bf 81}, 4798 (1998).
\bibitem{prl01} Y.M. Cho, Phys. Rev. Lett. {\bf 87}, 252001 (2001).
\bibitem{plb04} Y.M. Cho, Phys. Lett. {\bf B603}, 88 (2004).

\bibitem{ijmpa08} Y.M. Cho, B.S. Park, and P.M. Zhang, Int. J. Mod. Phys.
{\bf A23}, 267 (2008).
\bibitem{epjc17} Y.M. Cho, Kyoungtae Kimm, J.H. Yoon, and 
Pengming Zhang, Euro. Phys. J. {\bf C616}, 77 (2017).
\bibitem{ran} A.F. Ranada, Lett. Math. Phys. {\bf 18}, 97 (1898);
J. Phys. {\bf A 25}, 1621 (1992).
\bibitem{nat} W. Irvine and D. Bouwmeester, Nature Physics {\bf 4}, 
716 (2008).
\bibitem{knotprep} M. Arrayas, D. Bouwmeester, and J. Trueba,
Phys. Rep. {\bf 667}, 1 (2017).

\bibitem{traut} A. Trautman, Int. J. Theor. Phys. {\bf 16}, 561 (1977).  

\bibitem{wheel} J.A. Wheeler, Phys. Rev. {\bf 97},511 (1955).
See also K.W. Ford and J.A. Wheeler, {\it Geons, Black Holes, and 
Quantum Foam: A Life in Physics}, New York (W.W. Norton and Company)
2000. 

\bibitem{prd13} Y.M. Cho, Franklin H. Cho, and J.H. Yoon,
Phys. Rev. {\bf D 87}, 085025 (2013).
\bibitem{prd15} Y.M. Cho, X.Y. Pham, Pengming Zhang, Ju-Jun Xie, and
Liping Zou, Phys. Rev. {\bf D 91}, 114020 (2015).
\bibitem{atiyah} M. Atijah, {\it The Geometry and Physics of Knots,}
Cambridge University Press (1990).

\bibitem{berry1} M.V. Berry and M.R. Dennis, Proc. Roy. Soc. (London) 
{\bf A 457}, 2251 (2001); I. Bialynicki-Birula and Z. Bialynicki-Birula,
Phys. Rev. {\bf A 67}, 062114 (2003).
\bibitem{plb05} Y.M. Cho, Phys. Lett. {\bf B616}, 101 (2005).
\bibitem{moff} H.K. Moffat, J. Fluid Mech. {\bf 35}, 69 (1969).
\bibitem{berry2} M.V. Berry, Found. Phys. {\bf 31}, 659 (2001).
\bibitem{pra05} Y.M. Cho, Int. J. Pure Appl. Phys. {\bf 1}, 246 (2005); 
Y. M. Cho, Hyojoong Khim, and Pengming Zhang, Phys. Rev. {\bf A72}, 
063603 (2005).
\bibitem{berg} M.A. Berger, Plasma Phys. Control. Fusion {\bf 41},
b167 (1999).
\bibitem{kami} R.D. Kamien,  Rev. Mod. Phys. {\bf 74}, 953 (2002).
\bibitem{baba} E. Babaev, Phys. Rev. Lett. {\bf 88}, 177002 (2002);
E. Babaev, L. Faddeev, and A. Niemi, Phys. Rev. {\bf B65}, 100512 (2002).
\bibitem{prb06} Y.M. Cho and Pengming Zhang, 
Phys. Rev. {\bf B73}, 180506(R) (2006).
\bibitem{epjb08} Y.M. Cho and P.M. Zhang, Euro Phys. J. {\bf B65}, 155 (2008).

\bibitem{cqg13} Y.M. Cho, Franklin H. Cho, and J.H. Yoon, Class. Quantum
Grav. {\bf 30}, 055003 (2013).

\bibitem{prd80} Y.M. Cho, Phys. Rev. {\bf D21}, 1080 (1980).
See also Y.S. Duan and M.L. Ge, Sinica Sci. {\bf 11}, 1072 (1979). 
\bibitem{prl81} Y.M. Cho, Phys. Rev. Lett. {\bf 46}, 302 (1981); 
Y.M. Cho, Phys. Rev. {\bf D23}, 2415 (1981).
\bibitem{gies} S. Shabanov, Phys. Lett. {\bf B458}, 322 (1999);
H. Gies, Phys. Rev. {\bf D63}, 125023 (2001).
\bibitem{kondo} K. Kondo, Phys. Lett. {\bf B600}, 287 (2004);
K. Kondo, T. Murakami, and T. Shinohara, Euro Phys. J. {\bf C42},
475 (2005).
\bibitem{zucc} R. Zucchini, Int. J. Geom. Methods Mod. Phys.
{\bf 1} 813 (2004).

\bibitem{epjc15} Kyoungtae Kimm, J.H. Yoon, and Y.M. Cho, 
Eur. Phys. J. {\bf C75}, 67 (2015); Kyoungtae Kimm, 
J.H. Yoon, S.H. Oh, and Y.M. Cho, Mod. Phys. Lett. 
{\bf A31}, 1650053 (2016).
\bibitem{piet} B. Piette, B. Schroers, and W. Zakrzewski,
Nucl. Phys. {\bf 439}, 205 (1995).
\bibitem{plb07} Y.M. Cho, Phys. Lett. {\bf B644}, 208 (2007).
\bibitem{glad} J. Gladikowski and M. Hellmund,
Phys. Rev. {\bf D56}, 5194 (1997).

\bibitem{ussr} L. Kapitansky and A. Vakulenko, Sov. Phys. Doklady
{\bf 24}, 433 (1979); F. Lin and Y. Yang, Commun. Math. Phys.
{\bf 249}, 273 (2004).
\bibitem{batt} R. Battye and P. Sutcliffe, Proc. R. Soc.
Lond. {\bf A455}, 4305 (1999).
\bibitem{fadd2} L. Faddeev and A. Niemi, Phys. Rev. Lett.
{\bf 82}, 1624 (1999); Phys. Lett. {\bf B449}, 214 (1999);
{\bf B464}, 90 (1999).

\bibitem{ho} It has been well-known that the vorticity
plays a crucial role in $\rm ^3He$ superfluid.
See N. Mermin and T. Ho, Phys. Rev. Lett. {\bf 36}, 594 (1976);
G. Volovik, {\it The Universe in a Helium Droplet},
Clarendon Press (Oxford), 2003.
\bibitem{medal} B. Acharya et al. (MoEDAL Collaboration),
Phys. Rev. Lett. {\bf118}, 061801 (2017).

\end{thebibliography}
\end{document}